\documentclass[aps,prb,twocolumn,10pt,floatfix]{revtex4-2}
\usepackage[utf8]{inputenc}
\usepackage[T1]{fontenc}
\usepackage{lmodern}
\usepackage{amsmath}
\usepackage{amsfonts}
\usepackage{amssymb}
\usepackage{bm}
\usepackage{chemformula}
\usepackage{graphicx}
\usepackage{hyperref}
\usepackage{physics}
\usepackage{mathtools}
\usepackage{siunitx}
\usepackage{xcolor}
\usepackage{soul}
\newcommand{\bmr}{\textbf{r}}
\newcommand{\bmk}{\textbf{k}}

\newcommand{\tcm}{\textcolor{black}}

\makeatletter

\newcommand{\Rmnum}[1]{\expandafter\@slowromancap\romannumeral #1@}
\newcommand{\colorcaption}[2][]{
	\begingroup
	\renewcommand{\@caption@fignum@sep}{ (color online). }
	\caption[#1]{#2}
	\endgroup
}
\makeatother

\begin{document}
	\title{\textbf{Photon absorption in twisted bilayer graphene}}
	\author{Disha Arora, Deepanshu Aggarwal, Sankalpa Ghosh and Rohit Narula}
	\affiliation{Department of Physics, Indian Institute of Technology Delhi, New Delhi-110016, India}
	
	\begin{abstract}
		We investigate one- and two-photon absorption in twisted bilayer graphene (TBLG) by examining the effects of tuning the twist angle $ \theta $ and the excitation energy $ E_l $ on its absorption coefficients $ \alpha_{i=1,2}$. \tcm{We find that $ \alpha_1 $ as a function of $ E_l $ for TBLG exhibits distinct peaks corresponding to its van Hove singularities (vHs). \tcm{For small twist angles, such as $\theta \sim 1.8^{\circ}$, the magnitude of the resonant peak for $\alpha_1$ is roughly twice that of bilayer graphene (BLG). This \tcm{enhanced} response, compared to BLG, can be attributed to the increased density of states (DOS) in the twisted structure.} However, as the twist angle increases the magnitude of the resonant peak approaches that of two decoupled single-layer graphene (SLG) sheets.} \tcm{On the other hand, the two-photon absorption coefficient} $ \alpha_2 $ for TBLG at low twist angles displays \tcm{an} enhancement of \tcm{about} one order of magnitude compared to SLG at the energies corresponding to the resonant peak, as well as a \tcm{small but} notable increase relative to BLG. \tcm{As the twist angle decreases from $ 8^{\circ} $ to $ 2.5^{\circ} $, the resonant peak intensifies by three orders of magnitude.} Interestingly, as $\theta$ increases the resonant features exhibited by $\alpha_{i=1,2}$ \textit{vs.} $ E_l $ shift progressively from the infrared to the visible. On doping TBLG, both $\alpha_1 $ and $ \alpha_2 $ \textit{vs.} $ E_l $ remain essentially unchanged  but with a \tcm{slight} red-shift in their resonant peaks. Additionally, we explore various polarization configurations for two-photon absorption and determine the conditions under which $\alpha_2$ becomes extremal.
	\end{abstract}
	\maketitle
	
	\section{Introduction}
	\tcm{Twisted bilayer graphene (TBLG)~\cite{Berger2006, Hass2007, Zhao2011, Xie2011, Li2010, Miller2010, Luican2011} is a two-dimensional superlattice formed by \tcm{the} vertical stacking of two graphene monolayers, with each layer rotated relative to the other}. This misalignment between the two layers gives rise to a moir\'{e} pattern~\cite{Moirebook} with localized $ AA $ and $ AB $-rich regions that form a moir\'{e} lattice. The interlayer interaction between the misoriented layers significantly modifies the low energy band structure thus endowing it with distinctive electrical~\cite{Zou2018,Long2023} and optical properties~\cite{Wang2010,Patel2019,Alencar2018,Zou2013,Ni2009,Righi2011,Sato2012,Robinson2013,Wang2013}. By tuning the twist angle, the location of the \tcm{van Hove} singularities (vHs) can be varied~\cite{Andrei2020,Li2010,Deepanshu2023} without the introduction of defects, chemical doping or electrical gating. At small twist angles \textit{e.g.,} $ \theta \sim 1.05^{\circ}$, TBLG exhibits almost flat energy bands ~\cite{Ni2008, Santos2007, Morell2010, Shallcross2008, Shallcross2010, Bistritzer2011, Santos2012, Trambly2010}, thus enabling features like unconventional superconductivity~\cite{Chichinadze2020,Cao2018Nat1,Choi2019} and Mott insulating states~\cite{Cao2018Nat2,You2019,Roy2019,Hoi2018,Xu2018,Fang2016,Mayou2012}.
	
	\begin{figure}
		\centering
		\includegraphics[scale=0.25]{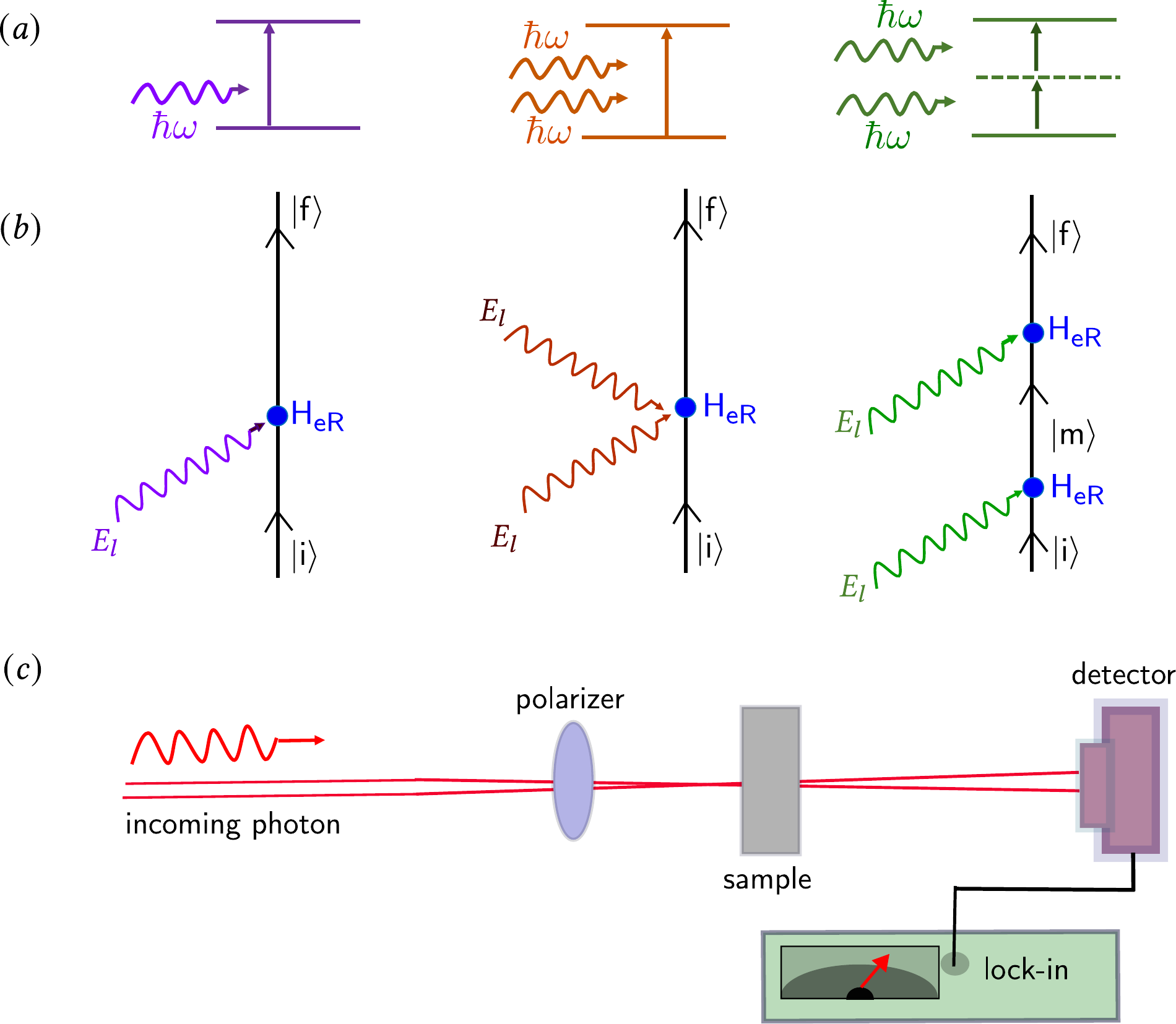}
		\caption{(a) \tcm{One-photon} absorption, two-photon absorption in the simultaneous and step-wise scheme, respectively, and (b) their corresponding Feynman diagrams. (c) \tcm{A schematic} of the experimental setup for \tcm{the process of absorption}~\cite{Krauss2021}.}
		\label{fig:feynman}
	\end{figure}
	Multiphoton absorption~\cite{Nathan85,Yee1971} (MPA) processes in crystalline solids have been the focus of much theoretical and experimental research~\cite{Gibson1976,Bass2008,Girlanda1971}. Particularly at short wavelengths, nonlinear absorption is a key factor in limiting the transparency of optical window materials and in producing laser-induced damage to optical components~\cite{Nathan85}. In general, these processes typically involve all of the solid's energy bands as intermediate states, so even a single MPA experiment can provide detailed information about the energy-band structure when combined with other pertinent experimental or theoretical data~\cite{Nair2008,Mak2008,Zhang2008,Li2009,Mak2010,Orlita2009} that is not easily obtained by using linear techniques. To induce a transition from the ground state to the excited state via the simultaneous absorption of two photons, the material is excited using a light source with a wavelength approximately twice as long as that required for one-photon absorption. Consequently, the rate of two-photon absorption (TPA) increases with the square of the photon flux.
In contrast, the rate of one-photon absorption (OPA) is directly proportional to the intensity ~\cite{Mariacristina2010}. Due to the low photon energy involved in TPA, the material under high \tcm{illumination} sustains less damage. TPA plays a crucial role in laser spectroscopy by enabling the transition between two states that cannot be coupled by an electric-dipole interaction~\cite{Braunstein1964,Grassano1992,Saissy1978}. It has been demonstrated~\cite{Sun2010} that quantum interference between OPA and TPA can solve the coherent control~\cite{Cui2015} and non-contact generation of ballistic photo-current in multilayer graphene, which have applications in quantum technology. TPA can also enable coherently excited states of molecules with energies in the far ultraviolet~\cite{Perez2003} when using visible light, for which coherent sources are conveniently available. Indeed, TPA provides a way of accessing a given excited state with the use of photons of half the energy (or twice the wavelength) of the corresponding one-photon transition, thus leading to applications in microscopy, micro-fabrication \cite{Maruo1997}, three-dimensional data storage \cite{Corredor2006,Dimitri1989}, optical power limiting \cite{Pawlicki2009}, two-photon fluorescence imaging~\cite{Denk1990} and sensing \cite{He2008}. Unlike all other nonlinear optical processes, such as second harmonic generation~\cite{Lee2008}, TPA has the benefit of generating a direct electrical response to the optical signal without the need for phase matching. As a result, it is also used for second-order autocorrelation, which lets one calculate optical logic operations and pulse duration~\cite{Piccardo2018}.
	
	\begin{figure}
		\centering
		\includegraphics[scale=0.37]{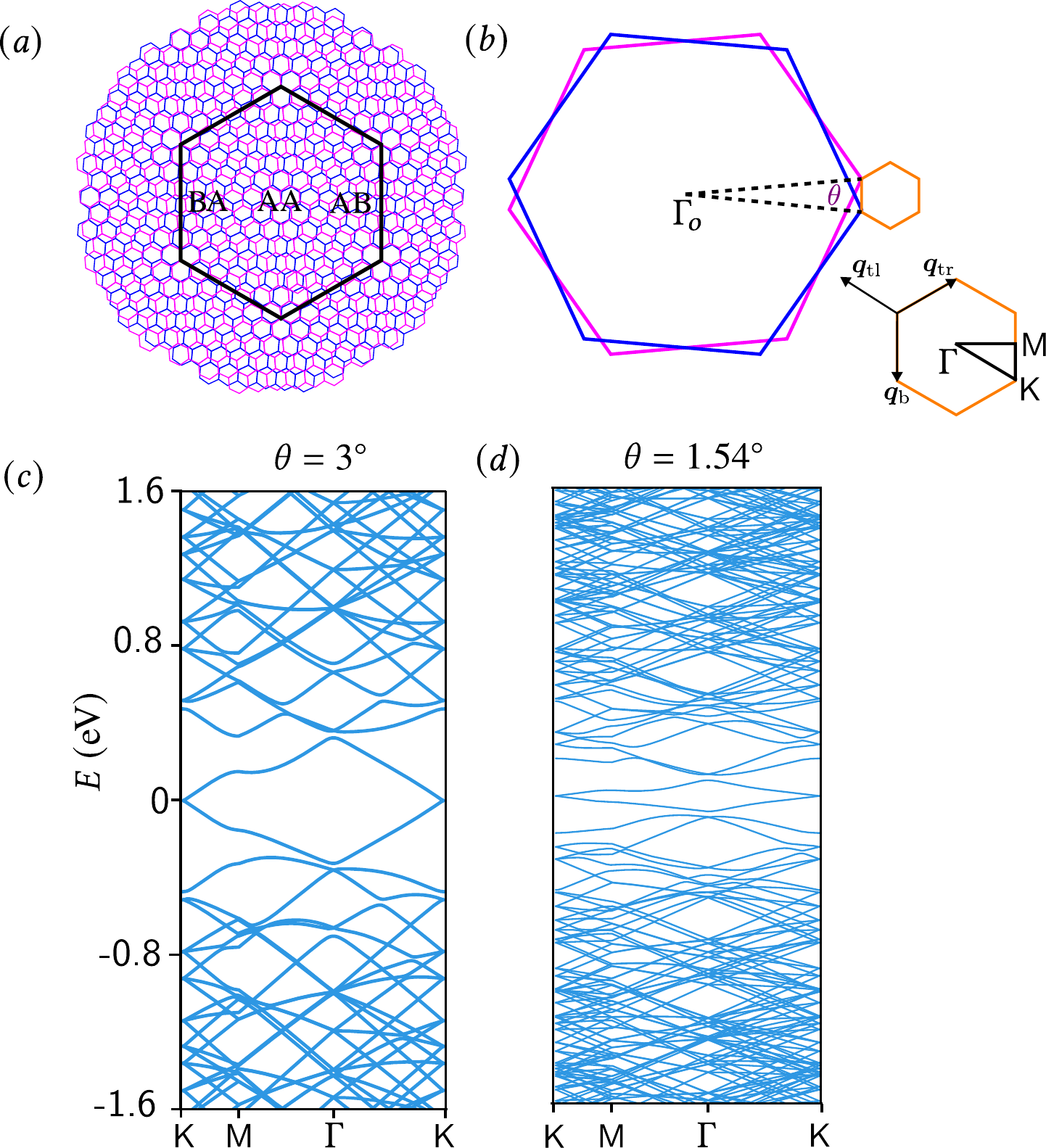}
		\caption{A two-dimensional view of TBLG in (a) real space and (b) reciprocal space, respectively. The small orange hexagon outlines the moir\'e Brillouin zone. The band structure along the high-symmetry path of the MBZ for twist angles (c) $ \theta = 3.00^{\circ} $ (d) $ \theta = 1.54^{\circ} $. The $w$ parameters are taken from Ref.~\cite{Koshino2018}.}
		\label{fig:lattice}
	\end{figure}
	
	The absorption of photons is an elementary process to all the higher-order processes involving light-matter interactions like Raman scattering~\cite{Hellwarth1963,Gardiner1989}, one-, two- or
	three-photon scattering~\cite{Fedorov1996,Nathan85}, \textit{etc}. In this paper, we study the polarization-controlled absorption characteristics of TBLG, specifically focusing on the absorption of one and two photons. We explore the effects of tuning the twist angle $ \theta $ and the excitation energy $ E_l$ on their corresponding absorption coefficients $ \alpha_{i=1,2}$. \tcm{Our findings demonstrate that $ \alpha_1 $ as a function of $ E_l $ for TBLG exhibits distinct peaks associated with the \tcm{latter's} vHs. For twist angles $\theta \lesssim 6.5^{\circ} $, the resonant peak of $ \alpha_1 $ demonstrates a \tcm{enhanced} response compared to BLG, which is attributed to the increased density of states (DOS). \emph{E.g.}, at $\theta = 2.5^{\circ}$, the magnitude of the resonant peak is enhanced by a factor of $ \sim 1.6$. However, as the twist angle increases the magnitude of the resonant peak of $ \alpha_1 $ converges toward that of two decoupled sheets of SLG~\cite{PhysRevB.78.113407,PhysRevLett.100.125504,PhysRevLett.101.056803,Uchida}}. Furthermore, the TPA coefficient $\alpha_2$ in low twist angle limit and at energies corresponding to the resonant peak, TBLG shows an enhancement of about one order of magnitude compared to SLG. \tcm{Interestingly, on increasing the twist angle between the two layers, the resonant features observed in the absorption coefficients progressively shift from the infrared region to the visible part of the energy spectrum, \tcm{which is \tcm{in accord} with the shifting of vHs to higher energies on increasing the twist angle. This ability to control the resonant feature of the absorption spectrum makes TBLG a highly tunable material, allowing researchers to optimize its optical absorption properties across a range of laser energies. This optimization is useful in designing devices such as photodetectors~\cite{ELCHAAR20112165,Dutta2005}, solar cells~\cite{asi5040067,PIPREK20033,PIPREK2003121}, and sensors that operate efficiently at \tcm{the} desired \tcm{optical frequencies}.} Furthermore, the gradual shift of the resonant peak from \tcm{the} infrared to visible energies offers a practical experimental method for determining the twist angle. By measuring the energy at which \tcm{the} maximum absorption occurs and linking it to the density of states (DOS), researchers can infer the twist angle, \tcm{thereby} enhancing the study of TBLG's properties and its potential applications in \tcm{the frontiers of} materials science.} Additionally, we delve into various polarization configurations for the two photons independently, determining the conditions under which $\alpha_2$ becomes extremal.
		
	This paper is organized as follows: Sec.~\ref{sec:theory} discusses the theory of the multiphoton absorption process and provides the relevant expressions used in our calculations. The interaction of photons with TBLG is discussed and derived in Sec.~\ref{sec:IntrotoTBG}. In section.~\ref{sec:results} we discuss the behavior of the OPA and TPA coefficients as a function of the incident photon energy. Our conclusions and scope of future work are outlined in Sec.~\ref{sec:conc}.
			\begin{figure*}
		\centering
		\includegraphics[scale=0.27]{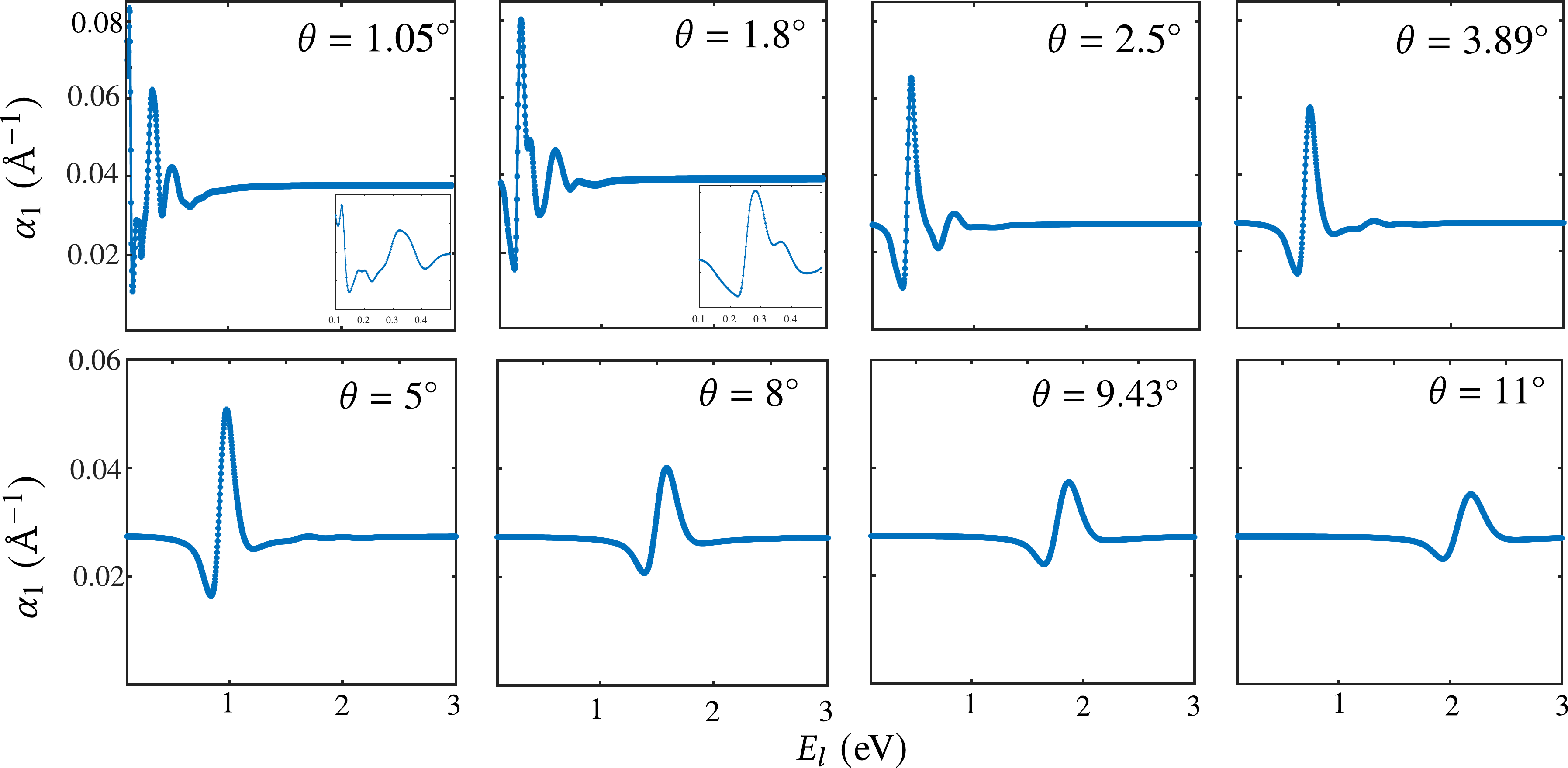}
		\caption{\tcm{The one-photon absorption coefficient ($ \alpha_1 $) as a function of the (X-polarized) incident excitation energy ($ E_l $) for various twist angles. The insets for $ \theta=1.05^{\circ}, 1.8^{\circ} $ show the details of the resonance feature at low $E_l$}.}
		\label{fig:R1}
	\end{figure*}
	\begin{figure}
		\centering
		\includegraphics[scale=0.4]{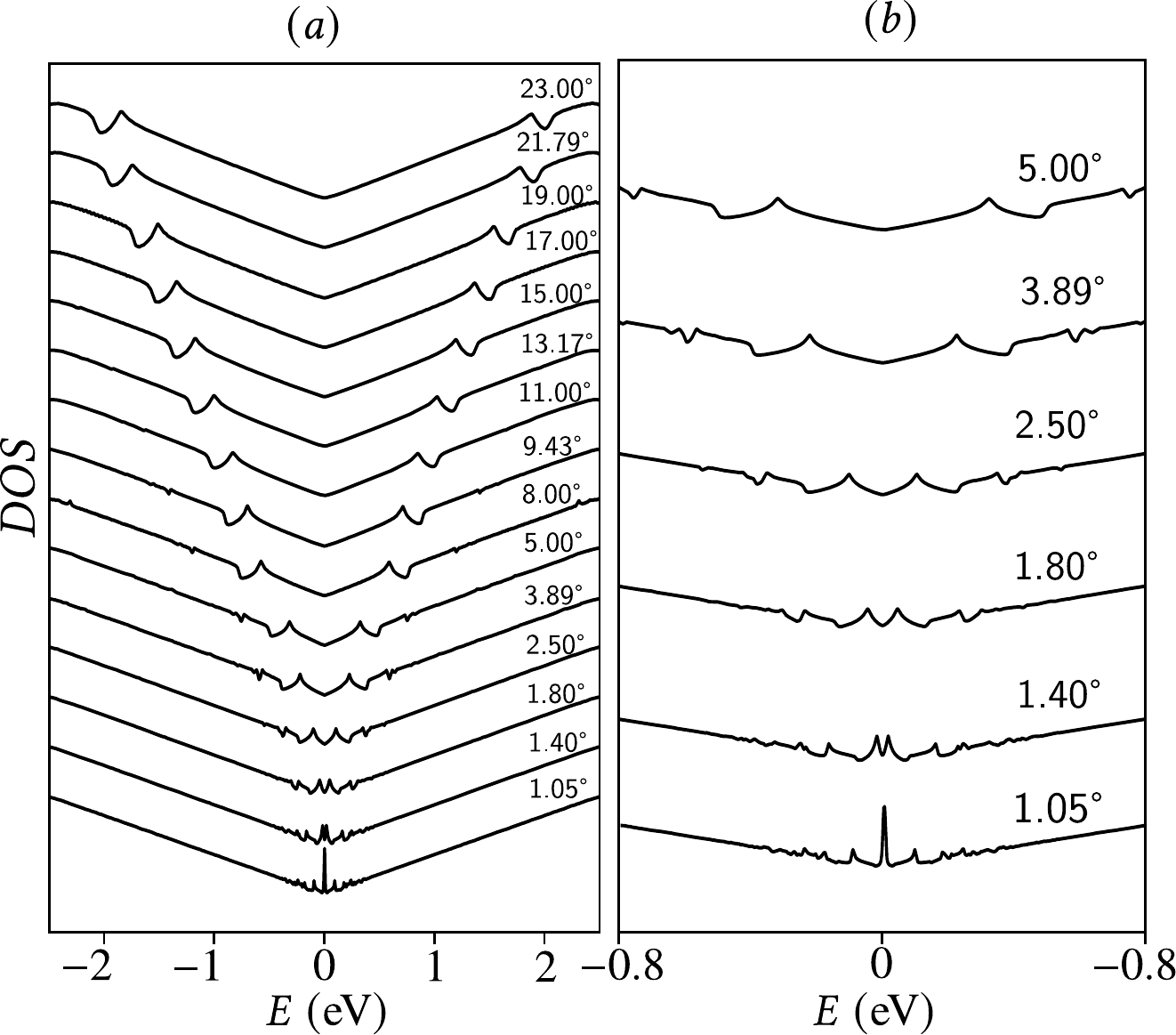}
		\caption{The density of states for twisted bilayer graphene at various twist angles for (a) a wide and (b) narrow energy range. The position of the vHs shifts to higher energy as the twist angle increases.}
		\label{fig:DOS}
	\end{figure}
	\section{Theoretical background}
	\label{sec:theory}
	
	Theoretically, the microscopic and macroscopic techniques have been used to study MPA~\cite{Nathan85}. In the former, the photon-induced electronic transition probabilities are used to compute the absorption coefficients, whereas, in the latter, they are derived from the imaginary part of the macroscopic current operator $\textbf{j}$~\cite{Malic2006,Haug2009}. In Appendix~\ref{sec:appendix1}, based on Refs~\cite{Wang2016,Gao2021}, we evaluate the macroscopic current operator and the absorption coefficients for TBLG. For our purposes, to calculate the absorption coefficients within the microscopic picture, we follow the approach developed by G\"{o}ppert-Mayer~\cite{Mayer2009} in which the $n^{th}$-order time-dependent perturbation theory is employed to derive an expression for the probability of the simultaneous absorption of $n$ photons by a single atomic electron. Applying this elementary idea to a crystalline material provides the probability of a direct electronic transition from an initial valence band $v$ to a final conduction band $c$, accompanied by the simultaneous absorption of $n$ photons, each of frequency $ \omega $ and is given as \cite{Nathan85,Cohen1992,Qiu2024}:

	\begin{equation}
		W_n = \frac{2\pi}{\hbar}\int_{\textbf{k}}\abs{\mathcal{T}}^2\delta\left[E_c(\textbf{k})-E_v(\textbf{k})-n\hbar\omega)\right]\frac{\dd^3\textbf{k}}{(2\pi)^3}
		\label{eq:ntprob}
	\end{equation}
	where,
	\begin{equation}
		\mathcal{T} = 
		\sum_{m...i}\frac{\bra{\psi_c}H_{eR}\ket{\psi_m}\bra{\psi_m}H_{eR}\ket{\psi_l}}{E_m - E_l-(n-1)\hbar\omega}...\frac{\bra{\psi_i}H_{eR}\ket{\psi_v}}{E_i-E_v-\hbar\omega}
	\end{equation}

	$ W_n $ is the $n^{th}$ order transition probability rate per unit volume. $ \psi_{i},\psi_{j}... $ are the Bloch functions of the crystalline electrons in bands $ i,j... $ with energies $ E_i,E_j... .$ respectively. Energy conservation is expressed by the delta function $ \delta $, and the summation is taken over all the possible intermediate states for a given transition. The summation over \textbf{k} is carried over the entire first Brillouin zone. $ H_{eR} = \frac{e}{c}\textbf{A}\cdot v_F\bm{\sigma} $ is the electron-radiation interaction Hamiltonian, where $ \textbf{A} = A\textbf{e}$ is the vector potential of the incident light wave of amplitude $ A $ and polarization vector $ \textbf{e} $. The \textit{n}-photon absorption coefficient is related to the $ n^{th} $-order transition probability as 
	\begin{equation}
		\alpha_n = g_sg_{\nu}\frac{2n\hbar \omega W_{n}}{I^{n}~d}
		\label{eq:alpha}
	\end{equation}
	
	where, $ I = \epsilon_o c \omega^2 A^2/2 $ is the incident radiation intensity (in $\si{\watt}/\si{\meter}^2$) \tcm{and the factors $g_s$ and $ g_{\nu} $ account for spin and valley degeneracy, respectively}~\cite{Brinkley2016}. It is usually challenging to calculate accurate numerical values from the MPA transition probabilities because they require the knowledge of the interaction Hamiltonian matrix elements among all the eigenstates and summation over all of its energy bands. Additionally, the knowledge of all the energy eigenvalues as functions of the wave vector \textbf{k} through the first Brillouin zone is also needed~\cite{Nathan85}.
		
		\begin{figure*}
	\centering
	\includegraphics[scale=0.34]{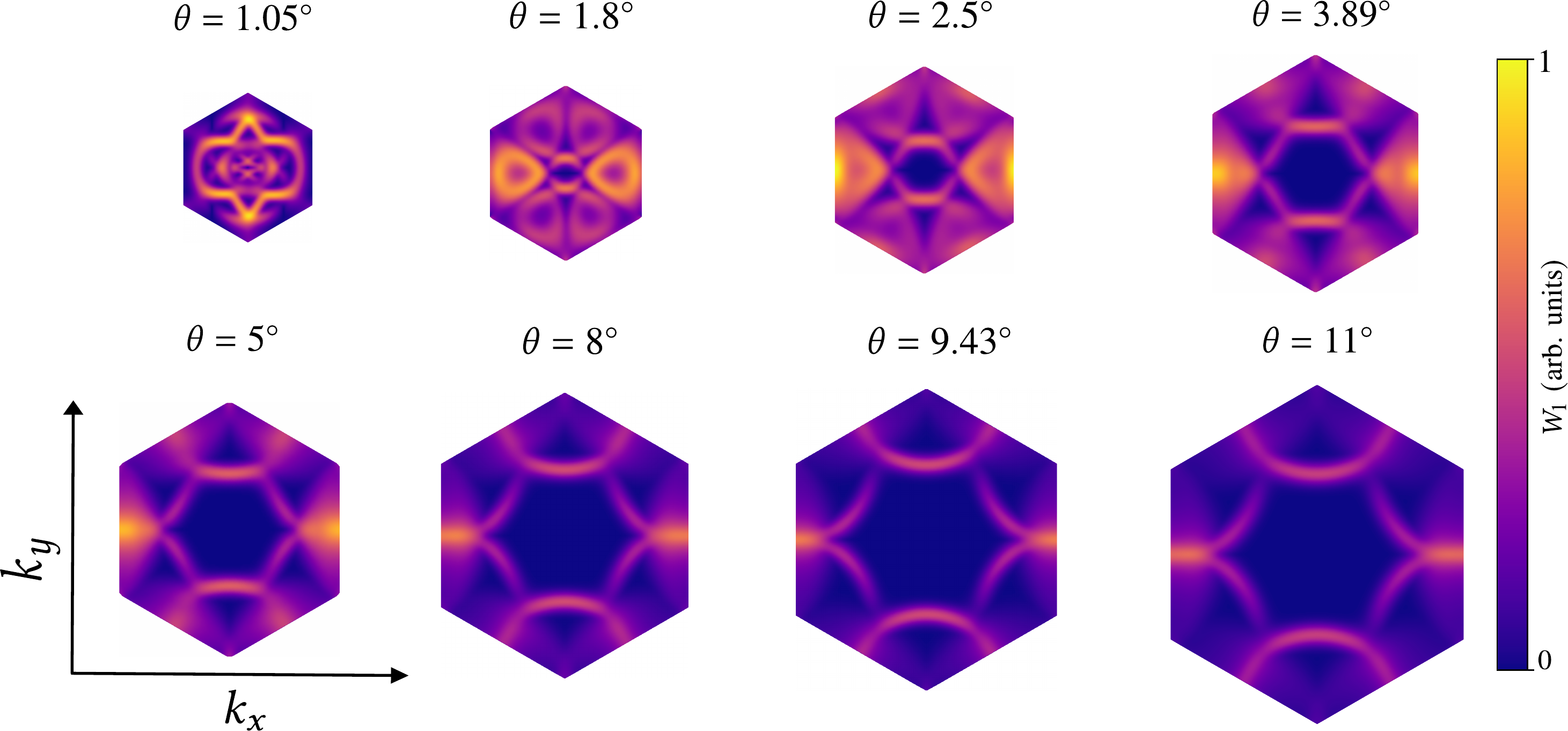}
	\caption{\tcm{Contour plots of the one-photon absorption transition probability at the energy corresponding to the resonant feature observed in Fig.~\ref{fig:R1}, shown as a function of $k_x$ and $k_y$ across the full moir\'e Brillouin zone for different twist angles (for $X$-polarized light, i.e., $\textbf{A} = (A_x,0)$, with $\textbf{A}$ being the vector potential). The brighter regions indicate the dominant contribution. The side length of the moir\'e Brillouin zone increases with the increase in twist angle.}}
	\label{fig:resonancex}
\end{figure*}

		\begin{figure*}
	\centering
	\includegraphics[scale=0.34]{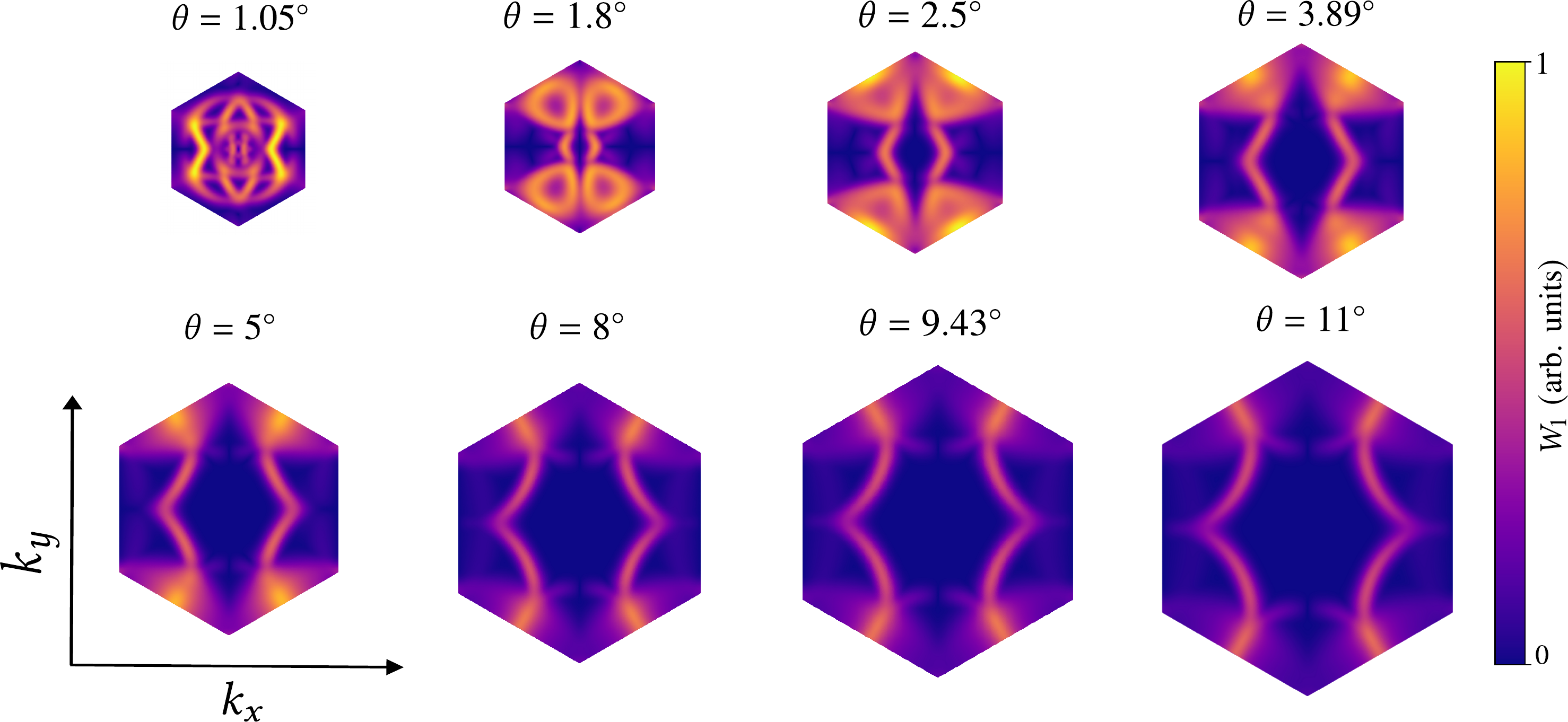}
	\caption{\tcm{Contour plots of the one-photon absorption transition probability at the energy corresponding to the resonant feature observed in Fig.~\ref{fig:R1}, shown as a function of $k_x$ and $k_y$ across the full moir\'e Brillouin zone for different twist angles (for $Y$-polarized light, i.e., $\textbf{A} = (0,A_y)$, with $\textbf{A}$ being the vector potential). The brighter regions indicate the dominant contribution. The side length of the moir\'e Brillouin zone increases with the increase in twist angle.}}
	\label{fig:resonancey}
\end{figure*}

\begin{figure*}
		\centering
		\includegraphics[scale=0.38]{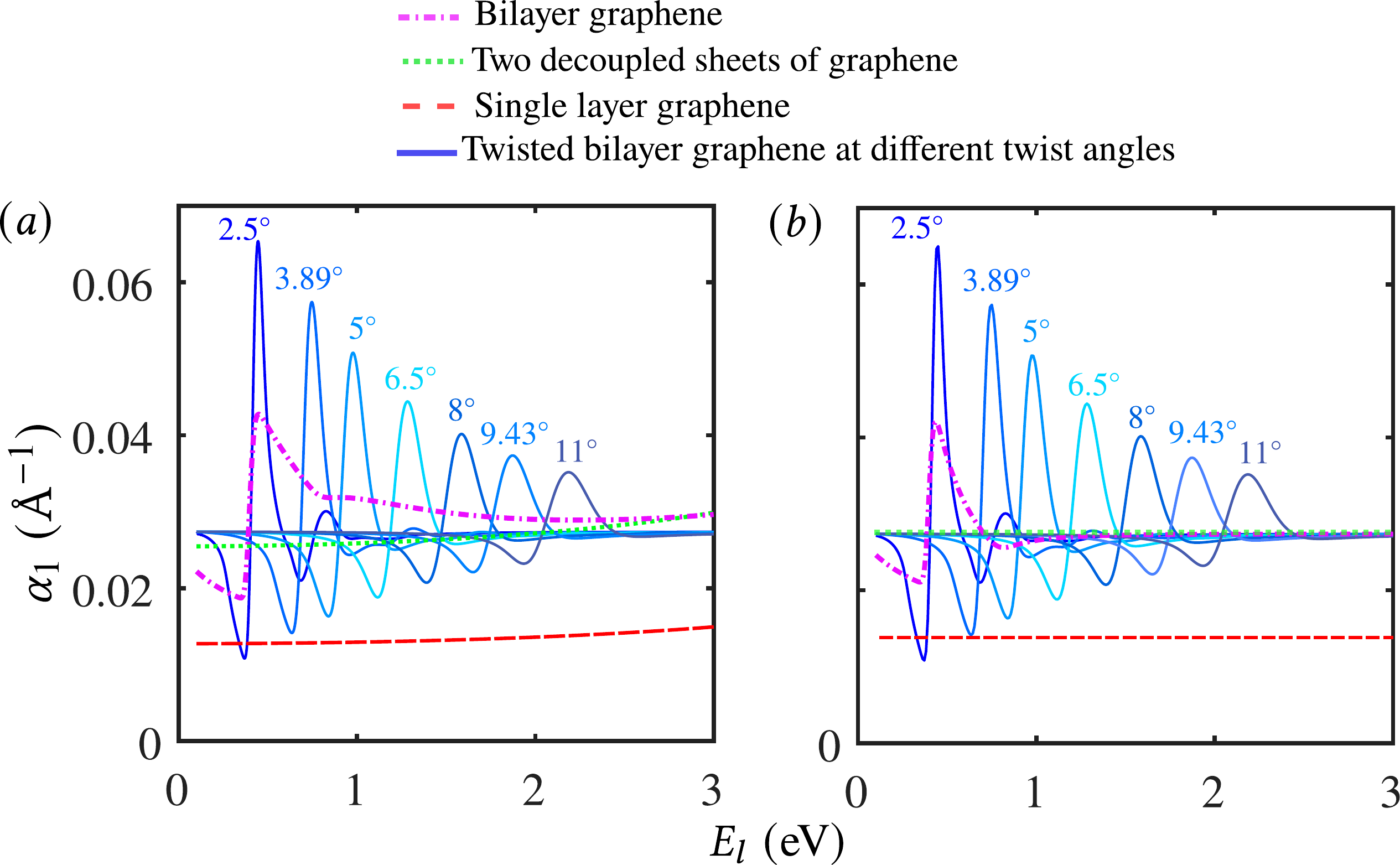}
		\caption{\tcm{The one-photon absorption coefficient, $\alpha_1$, is plotted as a function of incident excitation energy $E_l$. Twisted bilayer graphene is modeled using the BM model~\cite{Bistritzer2011}, while single-layer and bilayer graphene are described using (a) the first nearest-neighbor tight-binding approximation~\cite{McCann2013,Kadi2014} and (b) a continuum model~\cite{McCann2007}, respectively. The shaded region highlights the visible energy range.}}
		\label{fig:alpha1}
	\end{figure*}
	In this paper, we focus on the single- and two-photon absorption processes for TBLG by considering its full band structure while involving all \tcm{the possible transitions in a given energy range} between the occupied and unoccupied states. FIG.~\ref{fig:feynman}~(a) represents the schematic of the OPA process, where an electron transitions from a valence band to a conduction band state on the absorption of a single photon of frequency $ \omega $. FIG.~\ref{fig:feynman}~(b) shows the corresponding Feynman diagrams. The transition probability rate per unit volume for this process is given by
	\begin{equation}
		W_1= \frac{2\pi}{\hbar}\int_{\textbf{k}}\abs{\bra{f}H_{eR}\ket{i}}^2 \delta(E_f-E_i-\hbar\omega)\frac{\dd^3\textbf{k}}{(2\pi)^3}
	\end{equation}
	and the OPA coefficient is 
	\begin{equation}
		\alpha_1 = g_sg_{\nu}\frac{2\hbar\omega W_1}{I~d}
	\end{equation}
	
	TPA is an interband transition wherein two photons of energy $ \hbar\omega $ are absorbed simultaneously, creating an electron in the conduction band and a hole in the valence band~\cite{Mariacristina2010}. Both the photons can either be absorbed simultaneously or in a step-wise manner. FIG.~\ref{fig:feynman} shows that the leading contribution to the matrix element for the first process to TPA is of the first order in the interaction term. Both the photons are annihilated at the same point, which makes this process analogous to OPA. In contrast, the leading contributions of the second possible process are second-order in the interaction term. The significant difference between both processes is the lack of the presence of an intermediate state in the absorption involving the destruction of both photons simultaneously~\cite{Forbes2016}. 
	
	FIG.~\ref{fig:feynman}~(c) shows the schematic of an experimental setup for photon absorption. We have a laser source that emits photons of frequency $\omega$, a polarizer that controls the polarization direction of the incident light, the material system under consideration (TBLG in our case), and the photon detector. The transmitted signals are measured using the lock-in technique~\cite{Cosens1934,Michels1938,Michels1941}, which detects small AC signals obscured by a noisy environment. The photons produced by the laser source pass through the polarizer and interact with the electronic subsystem of the material via the light-matter interaction Hamiltonian $ H_{eR} $.	
	
	\tcm{The transition probability per unit volume for the TPA from an initial occupied to a final unoccupied state is given by~\cite{Nathan85}
		\begin{equation}
			W_2 = \frac{2\pi}{\hbar}\int_{\textbf{k}}\sum_{i,f}\abs{\mathcal{T}_{fi}}^2\delta\left[E_f(\textbf{k})-E_i(\textbf{k})-2\hbar\omega)\right]\frac{\dd^3\textbf{k}}{(2\pi)^3}
			\label{eq:transprob}
		\end{equation}
		where,
		\begin{equation}
			\mathcal{T}_{fi} = 
			\sum_{m}\frac{\bra{\psi_f}H_{eR}\ket{\psi_m}\bra{\psi_m}H_{eR}\ket{\psi_i}}{E_m - E_i-\hbar\omega}
			\label{eq:transitionmatrixdiag}
	\end{equation}}
	
	\tcm{and hence the expression for TPA coefficient becomes
		\begin{equation}
			\alpha_2 = g_sg_{\nu}\frac{4\hbar\omega~W_2}{I^2~d} 
	\end{equation}}

	\begin{figure}
	\centering
	\includegraphics[scale=0.34]{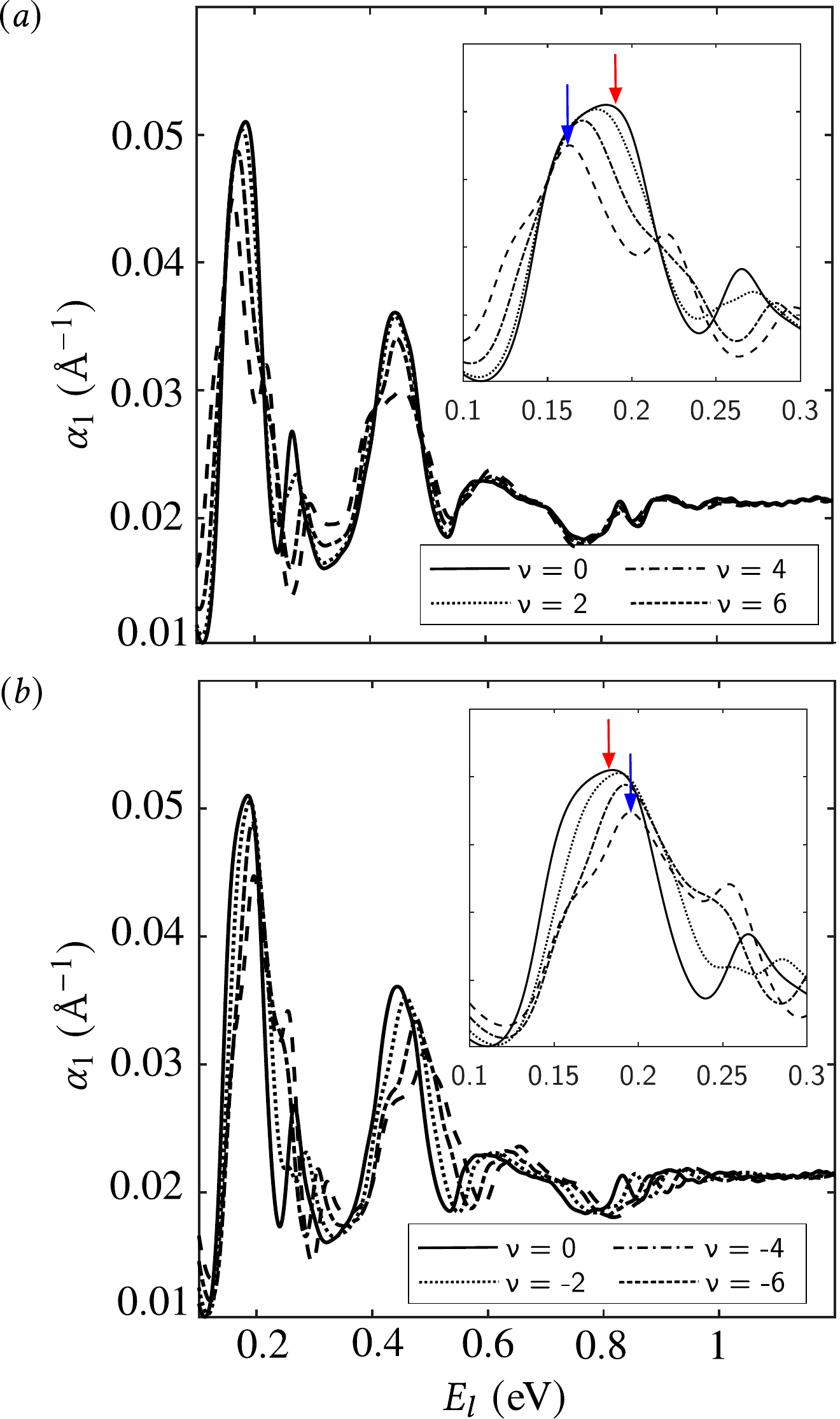}
	\caption{\tcm{The one-photon absorption coefficient as a function of the incident excitation energy for twist angle $ \theta = 1.54^{\circ}$ for (a) electron and (b) hole doping, respectively. The red arrow indicates the major absorption peak for an undoped TBLG, while the blue arrow marks the peak for doped TBLG.}}
	\label{fig:doping}
\end{figure}
	The domain of integration over \textbf{k} includes the entire Brillouin zone (BZ) of the crystalline material system under study, and the summation over $ i $ and $ f $ includes all the initial and final states of transition. $\ket{\psi_i}$ and $\ket{\psi_f}$ represent the initial and final state of the transition with energies $ E_i $ and $ E_f $, respectively. The summation over the states $ \ket{m} $ (with energies $ E_m $) includes all the possible intermediate states \tcm{and the delta function is approximated by \tcm{a} Gaussian representation}. \tcm{In} order to evaluate the quantities from Eqs.~\ref{eq:transitionmatrixdiag}-\ref{eq:transprob}, for a given pair of initial and final state of transition, we take the summation over all the possible intermediate states, represented by the index $ m $ with all the bands converged up to $ 10^{-2}~\si{\electronvolt} $  within a specific energy window. The cut-offs for this energy window's minimum and maximum energies depend on the value of the incident laser energy $ E_l $ such that the energy gap is far beyond $ E_l $ and is well converged. Under the assumption that the photon wave vector is negligible as compared to the electronic wave vector (dipole approximation) \cite{QMbook}, the vector potential $\textbf{A}$ can be separated from the expectation value. The term $ \bra{\psi_f}H_{eR}\ket{\psi_i} $ appearing in the numerator of Eq.~\ref{eq:transitionmatrixdiag} will be generically referred to as the "optical matrix element" in the subsequent discussion. 

	In the literature, there have been several experimental techniques like, nonlinear	transmission (NLT)~\cite{Ralston2003,Kleinman2003,Wenzel1973,PENZKOFER1976,PENZKOFER1976_1,Bechtel1976}, nonlinear luminescence (NLL)~\cite{Catalano1974,Johnston1980}, nonlinear photoconductivity (NLP)~\cite{Catalano1972,AMiller1979}, or the Z-scan technique~\cite{Sheik1990,inproceedings2013} to determine the value of the OPA and TPA absorption coefficients. In the next section, we shall discuss the continuum model of TBLG~\cite{Bistritzer2011} and its interaction with photons.
	\section{The model Hamiltonian}\label{sec:IntrotoTBG}
	FIG. \ref{fig:lattice}~(a) represents a two-dimensional (top) view of TBLG where the two graphene layers are rotated by a misorientation angle $ 
	\theta $ w.r.t. each other. The black hexagon marks the moir\'{e} unit cell with the $ AA $ and $ AB $ regions indicated. The corresponding rotated Brillouin zones are depicted in FIG.~\ref{fig:lattice} (b), where the vectors $ \textbf{q}_{\text{b}} $, $\textbf{q}_{\text{tr}}$, and $\textbf{q}_{\text{tl}}$, respectively, denote the Dirac points of the top and bottom layers. They stand for the momentum transfer associated with the three interlayer hopping processes marked by the black arrows in the MBZ. FIG.~\ref{fig:lattice}~(c) and (d) shows the band structure along the high-symmetry path of MBZ. The number of bands in a given energy range increases as the twist angle $ \theta $ is reduced~\cite{Deepanshu2023}, accounting for increased transitions between the valence and conduction bands. By turning on the light-matter interaction, the operator $\bm{\nabla}$ is replaced by $ \bm{\nabla}\rightarrow \bm{\nabla}+ie\textbf{A}/\hbar $~\cite{Bistritzer2011_1, Wang2012, Chang2022}, where $ \textbf{A} $  is the vector potential of the incident light with a magnitude $ A $ and polarization vector $ \bm{e} $. The real-space \tcm{effective continuum} Hamiltonian for TBLG \tcm{including the effect of} \tcm{lattice relaxations} through  corrugation, doping-dependent Hartree interactions and \tcm{eh}-asymmetry is ~\cite{Bistritzer2011,Koshino2018,Tarnopolsky2019,Bernevig2021,Vafek2023}, 
	\begin{widetext}
		\begin{equation}
			H_{\text{TBLG}}=\sum_{\xi=\pm}\sum_{\ell=1,2}\int \dd[2]{\bmr}\hat{\Psi}^{(\ell)\dagger}_{\xi}(\bmr)
			\bigg[v_{F}\bm{\sigma}^{\xi}_{\ell}\cdot\left(-i\hbar\bm{\nabla} + e\textbf{A}\right) + V_{H,\xi}(\textbf{r})\bigg]\hat{\Psi}^{(\ell)}_{\xi}(\bmr)
			+ \int \dd[2]{\bmr} \left[\hat{\Psi}^{(1)\dagger}_{\xi}(\bmr) T'_{\xi}(\bmr) \hat{\Psi}^{(2)}_{\xi}(\bmr) + \text{h.c.} \right]
			\label{mat:rham}
		\end{equation}
	\end{widetext}
\begin{figure*}
		\centering
		\includegraphics[scale=0.25]{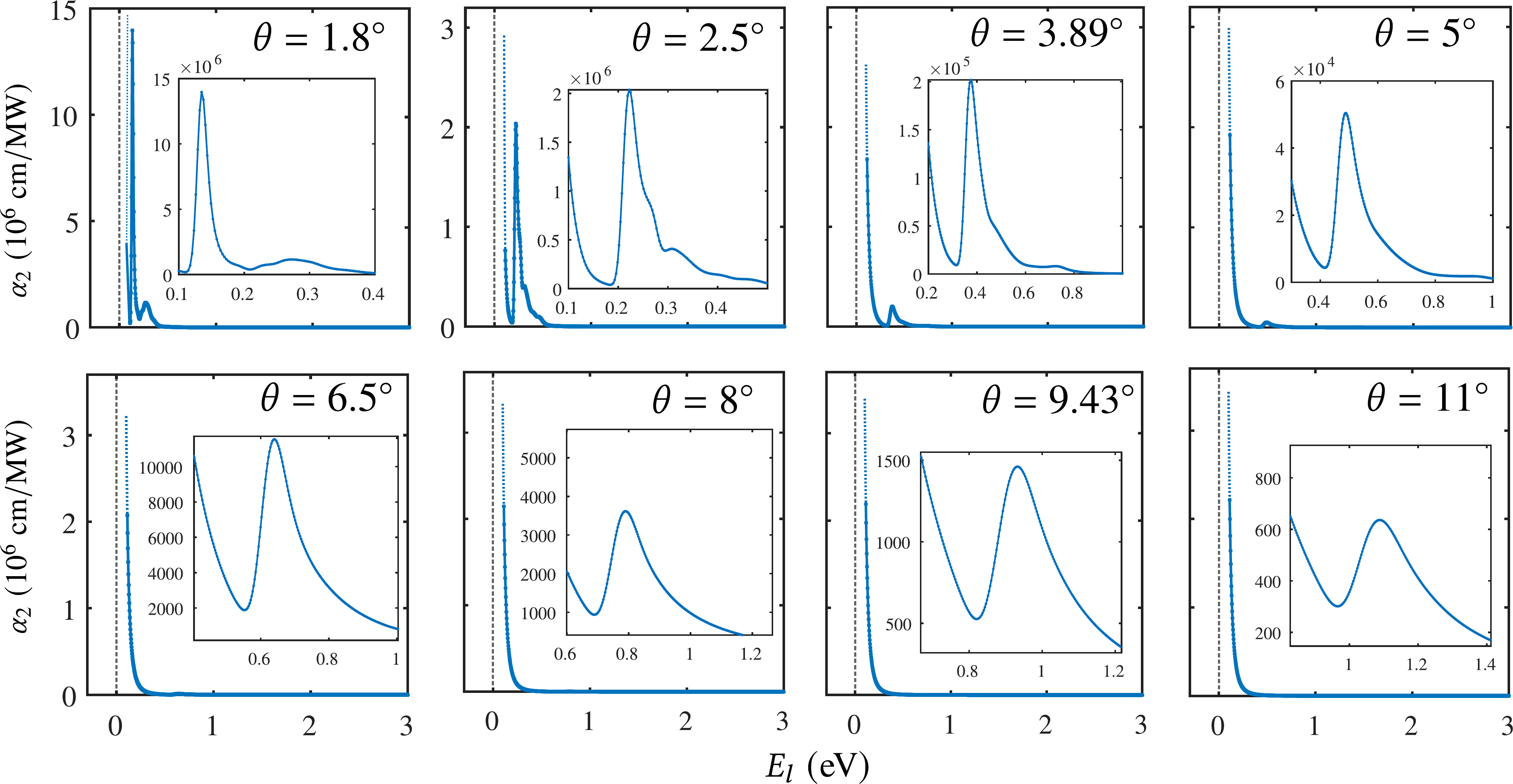}
		\caption{\tcm{The two-photon absorption coefficient $ \alpha_2 $ as a function of the ($X$-polarized) incident excitation energy $ E_l $. The inset of the figure shows the details of the resonance feature. The magnitude of the dominant absorption peak increases as the twist angle is reduced.}}
		\label{fig:2PA}
	\end{figure*}

	Where, $\hat{\Psi}^{(\ell)}_{\xi}(\bmr)$ is the two-component \tcm{fermion} field-operator corresponding to the layer $ \ell $, \tcm{valley $\xi$}, $ v_{F} $ is the Fermi velocity in graphene and  $ \bm{\sigma}^{\xi}_{\ell}(\theta) = e^{i\xi(-1)^{l}\sigma_z~\theta/2}\left(\xi\sigma_x,\sigma_y\right)e^{-i\xi(-1)^{l}\sigma_z~\theta/2}$ are the Pauli-matrices vector which accounts for the rotation of the individual graphene layer. On including the electron-electron interactions \emph{via} self-consistent Hartree calculations~\cite{Goodwin2020,Rademaker2019,Guinea2019,Guinea2022,Guinea2018}, it is found that the electronic band structure of a doped TBLG changes significantly. We employ the parametrization provided by Goodwin \emph{et al.}~\cite{Goodwin2020}, according to whom the doping and twist angle-dependent Hartree potential energy is described by
	$V_H(\textbf{r})\approx \mathcal{V}_{\theta}\sum_{j=1}^3~\cos(\bm{G}_j.\textbf{r})$
	where, $ \mathcal{V}_{\theta} = V(\theta)\left[\nu-\nu_o(\theta)\right] $. The quantity $ \nu_o(\theta) $ represents the doping level at which the Hartree potential vanishes, $ V(\theta)$ is a $ \theta $ dependent energy parameter, $ \nu $ is the \tcm{band} filling factor, and $ \bm{G}_j $ denotes the reciprocal lattice vectors that are used to describe the out-of-plane coordinates of TBLG~\cite{Koshino2018}. This equation resembles the continuum model provided in Ref.~\cite{Guinea2019}. The spatially dependent interlayer tunneling matrices $ T'(\textbf{r}) $ appearing as off-diagonal elements in Eq.~\ref{mat:rham} forms a
	smooth moir\'{e} potential~\cite{Bistritzer2011} of the form~\cite{Jung2014,Bernevig2021,Vafek2023}
	\begin{equation}
		T'(\textbf{r}) = \sum_{j=\text{b,tr,tl}} e^{-i\textbf{q}_j\cdot \textbf{r}}T'_j
		\label{eq:tmatexp}
	\end{equation}
	where, the $ T'_j $ matrices (for $\xi = 1$) are given by~\cite{Jung2014,Bernevig2021,Vafek2023}
	\begin{equation}
		T'_j = \left(w_o\sigma_o + iw_3\sigma_z \right)+  w_1\left[\sigma_x \cos\left(\phi^{'}\right) + \sigma_y \sin\left(\phi^{'}\right) \right]
		\label{eq:tmatVafek}
	\end{equation}
	\begin{figure*}
		\centering
		\includegraphics[scale=0.52]{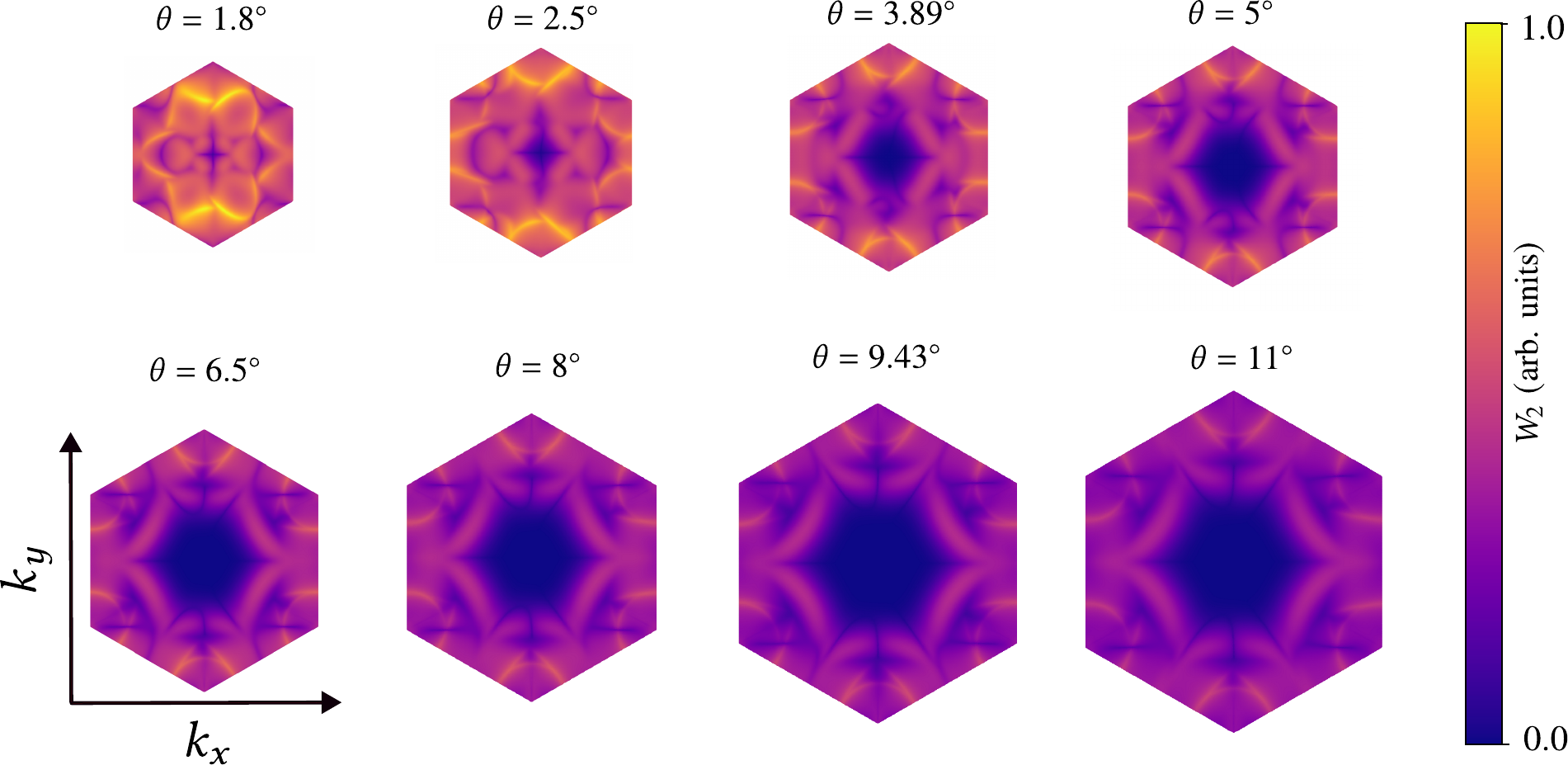}
		\caption{\tcm{Contour plots of the two-photon absorption transition probability at the energy corresponding to the resonant feature observed in Fig.~\ref{fig:2PA}, shown as a function of $k_x$ and $k_y$ across the full moir\'e Brillouin zone for different twist angles. The vector potential $\textbf{A}$ is along the $X$-direction for both the incident photons. The brighter regions indicate the dominant contribution. The side length of the moir\'e Brillouin zone increases with the increase in twist angle.}}
		\label{fig:2PAresonaceXX}
	\end{figure*}

    	\begin{figure*}
		\centering
		\includegraphics[scale=0.52]{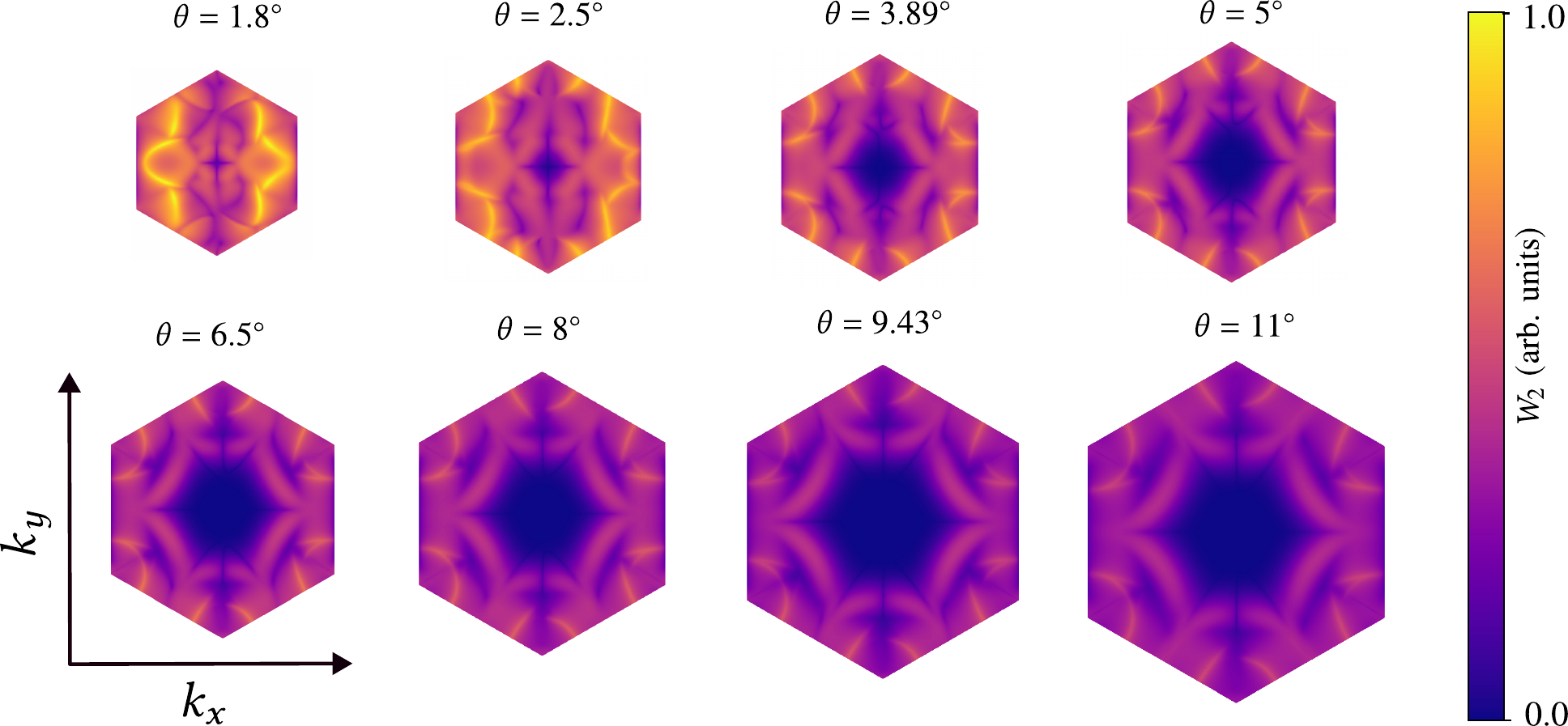}
		\caption{\tcm{Contour plots of the two-photon absorption transition probability at the energy corresponding to the resonant feature observed in Fig.~\ref{fig:2PA}, shown as a function of $k_x$ and $k_y$ across the full moir\'e Brillouin zone for different twist angles. The vector potential $\textbf{A}$ is along the $Y$-direction for both the incident photons. The brighter regions indicate the dominant contribution. The side length of the moir\'e Brillouin zone increases with the increase in twist angle.}}
		\label{fig:2PAresonaceYY}
	\end{figure*}

        \begin{figure*}
		\centering
		\includegraphics[scale=0.47]{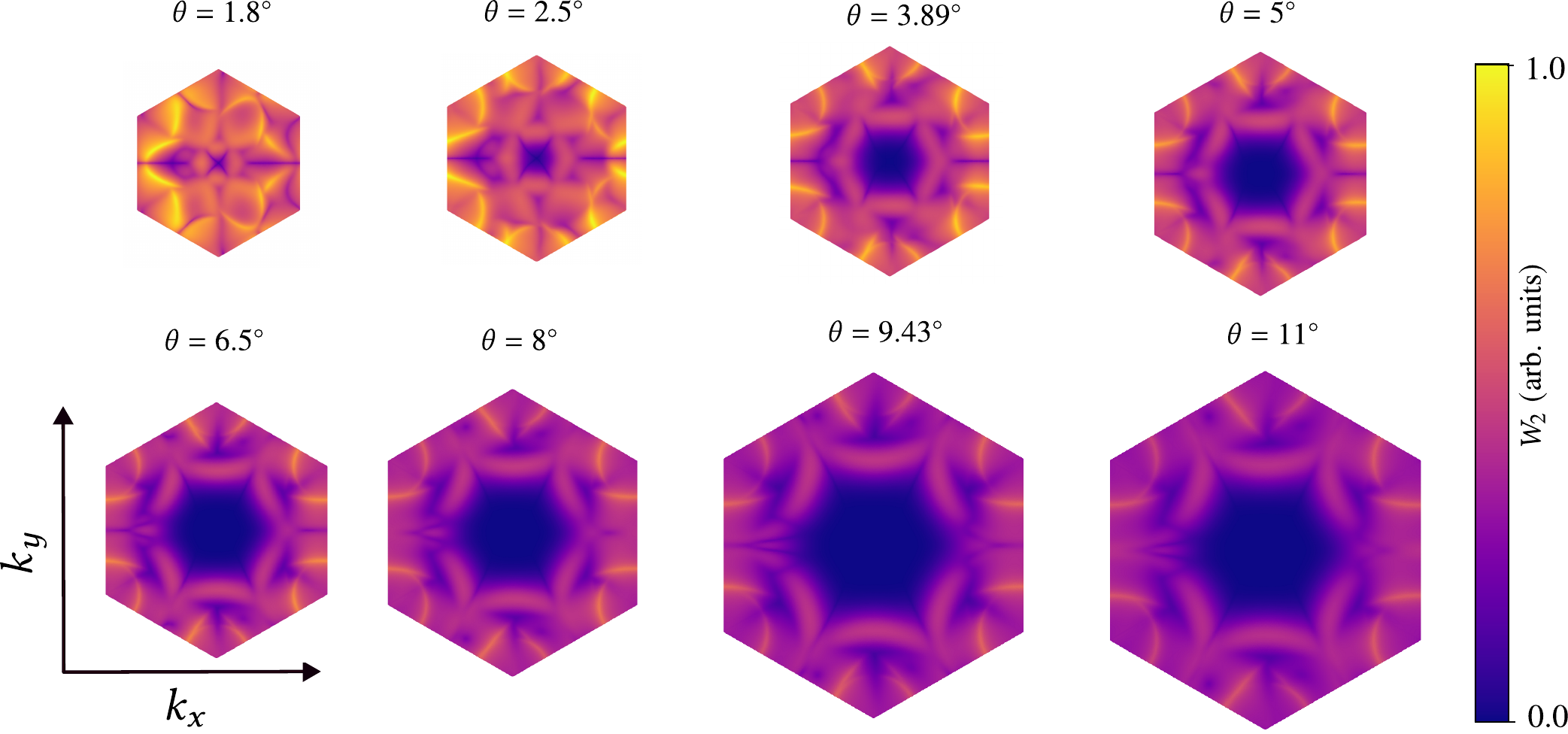}
		\caption{\tcm{Contour plots of the two-photon absorption transition probability at the energy corresponding to the resonant feature observed in Fig.~\ref{fig:2PA}, shown as a function of $k_x$ and $k_y$ across the full moir\'e Brillouin zone for different twist angles. The vector potential $\textbf{A}$ is along the $X$ and $Y$-direction for the two incident photons. The brighter regions indicate the dominant contribution. The side length of the moir\'e Brillouin zone increases with the increase in twist angle.}}
		\label{fig:2PAresonaceXY}
	\end{figure*}
    
	Here, $\phi^{'}=2\pi(j-1)/3$ and $ \bm{\sigma}$ are the Pauli matrices. The $ w_{o} $ term contributes to the diagonal elements. It represents the interlayer coupling between the A(B) sublattice of the top layer and the A(B) sublattice of the bottom layer. 
	The $ w_{1} $ term only contributes to the off-diagonal elements. It is thus associated with the interlayer coupling between the A(B) sublattice of the top layer and the B(A) sublattice of the bottom layer. $ w_3 $ is defined as the interlayer contact coupling, and as shown by Kang \emph{et al.}~\cite{Vafek2023}, it accounts for the dominant source of the non-negligible ph asymmetry in the model of Ref.~\cite{Fang2016}. Setting $ w_o = w_1$, $ w_3 = 0 $ and keeping
	$ \textbf{q}_{\text{j}} $ in the first shell recovers the tunneling matrices of the original BM continuum model~\cite{Bistritzer2011}. A minor contribution arises from the gradient coupling $ \lambda $, which has not been considered for our calculations. The corrugation effect substantially alters the electron band structure near the $ \Gamma $ point of the MBZ and can be included by setting $ w_o \neq w_1 $ in Eq.~\Ref{eq:tmatVafek}~\cite{Koshino2018,Carr2019}.
	
	The interaction of photons with TBLG is discussed and derived in our previous work~\cite{Arora2023}, where we also demonstrated that the optimum model used to describe the absorption of TBLG is to include corrugation effects~\cite{Koshino2018} in the continuum model of pristine TBLG~\cite{Bistritzer2011}. A similar demonstration is provided in Appendix~\ref{sec:appendix} for an arbitrary twist angle.
	
	\section{Results}
	\begin{figure*}
		\centering
		\includegraphics[scale=0.4]{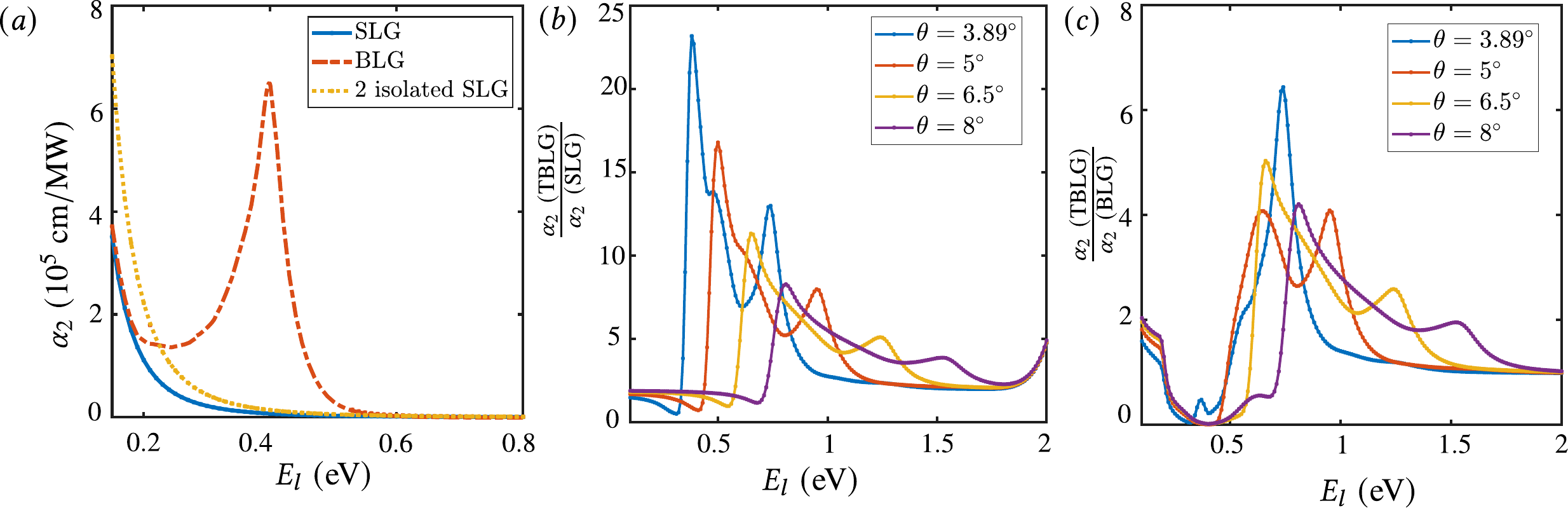}
		\caption{(a) The two-photon absorption coefficient $ \alpha_2 $ as a function of $ E_l$ for SLG, \tcm{two decoupled }sheets of graphene, and BLG. The TPA coefficient $ \alpha_2  $ of TBLG \tcm{as a ratio against} (b) SLG and (c) BLG, respectively.}
		\label{fig:TPAslgblgtbg}
	\end{figure*}
	\label{sec:results}
	\tcm{In the absence of the interlayer interaction in TBLG,
	the continuum Hamiltonian consists of two Dirac Hamiltonians due to individual layers, which leads to two
	uncoupled differential equations corresponding to each
	layer. Thus, the absorption coefficient
	can be calculated as
	\begin{equation}
		\alpha_{\text{uncoupled TBLG}} = g_sg_{\nu}g_{l} \alpha_{\text{graphene}}
	\end{equation}
	where $g_s = 2$ is spin-degeneracy factor, $g_{\nu} = 2$ is the
	valley/isospin-degeneracy factor, and $g_l = 2$ is the layer-degeneracy factor.}
	
	\tcm{In this section, we present our results for the behavior of $ \alpha_1 $ and $ \alpha_2 $ as a function of $ E_l $ obtained numerically for TBLG and provide a clear distinction with SLG and BLG.}
	
	\subsection{The one-photon absorption coefficient as a function of the incident laser energy}
    \tcm{By employing the BM model~\cite{Bistritzer2011}, we plot the numerically evaluated one-photon absorption coefficient, $ \alpha_1 $ as a function of excitation energy $ E_l $ for TBLG with twist angles ranging from $ 1.05^{\circ} $ to $ 11^{\circ} $ in FIG.~\ref{fig:R1}. The calculation considers summation over all the possible transitions from a given occupied state to an empty state. For higher twist angles, like $ \theta\gtrsim 2.5^{\circ} $, only one dominant absorption peak is present with its location associated with the vHs. As the twist angle decreases, apart from this resonant absorption peak, we see the presence of other minor peaks associated with the interaction of other peaks originating in the DOS (plotted in FIG.~\ref{fig:DOS}). The energy corresponding to the dominant absorption peak continuously decreases as the twist angle is reduced, accompanied by additional peaks in low energy regime, \emph{i.e.}, $ E_l \lesssim 1~\si{\electronvolt} $. For instance, the resonant feature in $ \alpha_1 $ for $\theta \sim 2.5^{\circ}$ occurs at $\sim 0.44~\si{\electronvolt}$. This energy value increases to $ \sim0.9~\si{\electronvolt} $ and $ \sim2.2~\si{\electronvolt} $ as $\theta$ is increased to $ \sim 5^{\circ} $ and $\sim 11^{\circ}$, respectively.}
    
    \tcm{To identify the region in the MBZ that gives the dominant contribution to $ \alpha_1 $, we provide the contour plots for the transition probability corresponding to the one-photon absorption as a function of $ k_x $ and $ k_y $ in the entire first MBZ of TBLG in FIG.~\ref{fig:resonancex} and ~\ref{fig:resonancey}, for the incident photon's polarization direction set along $ X $ and $ Y $, respectively. The excitation energies at which the contour is plotted is the energy corresponding to the major absorption peak in FIG.~\ref{fig:R1}. We see that as the twist angle increases, the location of the wave vectors giving the dominant contribution to OPA saturates.} 
    
    \tcm{Fig.~\ref{fig:alpha1}~(a) and (b)  provide a comparison of the numerically evaluated $ \alpha_1 $ as a function of $ E_l $ for TBLG (using the BM model~\cite{Bistritzer2011}) with SLG, two uncoupled sheets of graphene, and BLG in both the tight-binding description and continuum model, respectively. In the tight-binding description, $ \alpha_1 $ for SLG increases with the excitation energy, whereas in the continuum model, $ \alpha_1 = \frac{e^2}{2\hbar c \epsilon_o d} $, meaning it remains independent of the laser energy. We find that $ \alpha_1 $ for two uncoupled graphene sheets is twice the value for SLG. For BLG, we see an additional feature near $\hbar\omega \sim 0.4~\si{\electronvolt} $ or $ \lambda\sim 3100~\si{\nano\meter} $. This is in accord with the results presented in Refs.~\cite{Yang2011,Zhou2022}. This energy \tcm{corresponds} to the strongest inter-layer coupling between the pairs of dimer sites~\cite{McCann2013}. The additional feature \tcm{appearing in the vicinity of $ \hbar\omega \sim 0.4~\si{\electronvolt} $} in BLG arises due to the resonance between the pair of a low-energy band touching at the $K$ point and split bands separated by a band gap~\cite{McCann2013,McCann2007}. The continuum model saturates for higher photon energies, $ \hbar\omega\geq 2\gamma_1 $. This is in agreement with the analytical result provided in Ref.~\cite{McCann2007}.}

	\tcm{TBLG exhibits behavior similar to BLG, showing a characteristic peak in the absorption coefficient at the energy corresponding to the vHs, followed by saturation at higher energies. Our calculations show that for $ \theta \sim 2.5^{\circ} $, the magnitude of the resonant peak shows a significant enhancement, increasing by a factor of approximately 1.6 when compared to BLG. This suggests a notable increase in resonance intensity due to the rise in DOS. However, when the twist angle is increased to around $ 6.5^{\circ} $, the enhancement factor decreases to approximately 1.05. This indicates that the resonance intensity is only marginally higher than that of BLG, implying a weakening of interlayer coupling at larger twist angles. As the twist angle is further increased, the magnitude of the resonant peak gradually approaches that of two decoupled graphene sheets. This trend aligns with the behavior of other response functions, such as polarizability, as seen in Refs.~\cite{Stauber2016,PhysRevB.100.235424}.}
	
	\tcm{Furthermore, our results indicate that the resonance feature of $ \alpha_1 $ can be tuned to different excitation energies by varying the twist angle. For instance, when $ \theta \lesssim 9^{\circ} $, the peak of $ \alpha_1 $ occurs in the infrared region, whereas at higher twist angles, the resonance shifts progressively toward the visible range of excitation energies.}
	
	\begin{figure}
		\centering
		\includegraphics[scale=0.36]{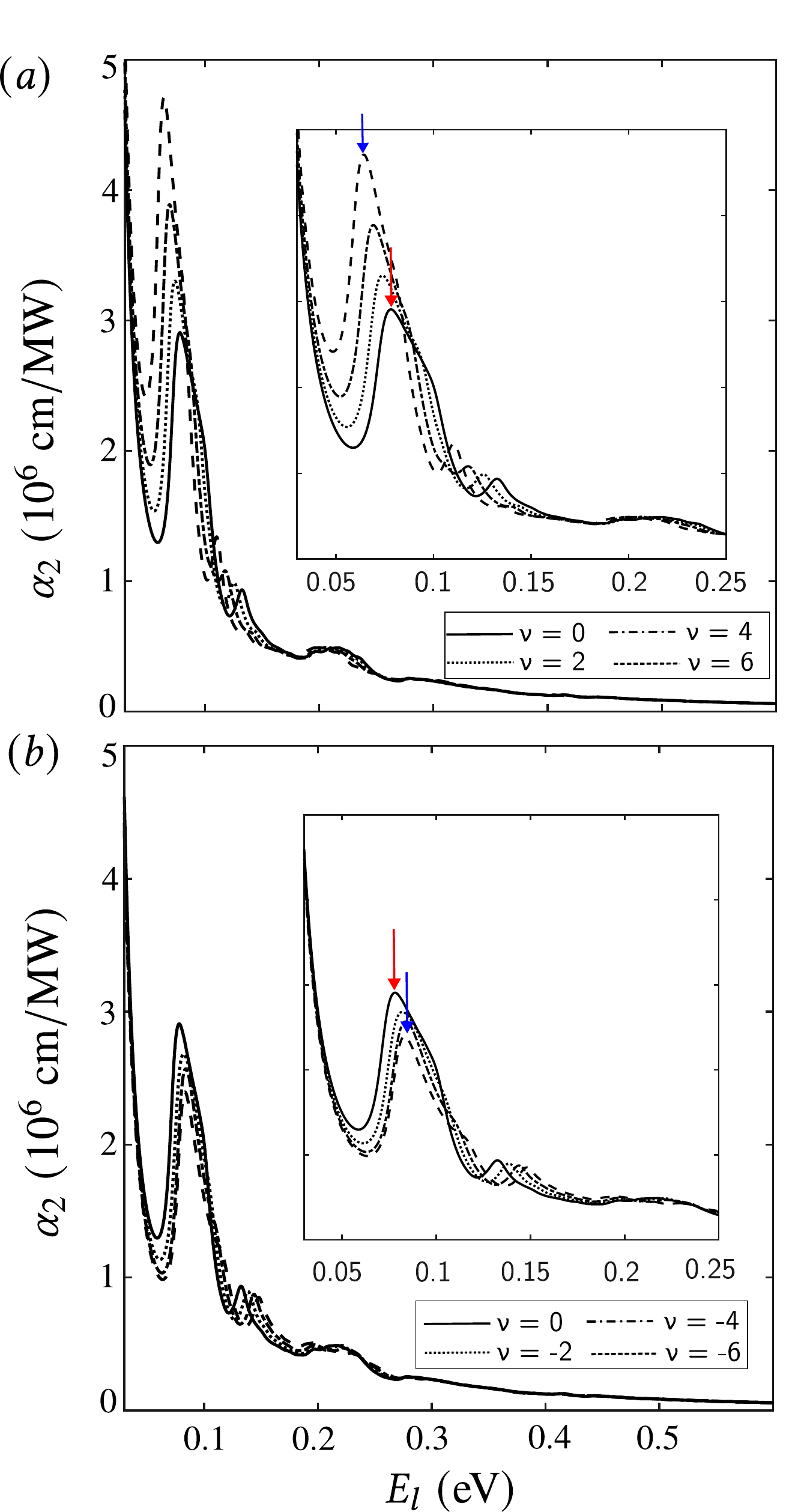}
		\caption{\tcm{The TPA coefficient $ \alpha_2 $ as a function of $ E_l $ for a twist angle $ \theta = 1.54^{\circ}$ for (a) electron and (b) hole doping, respectively. The red arrow indicates the dominant absorption peak for an undoped TBLG, while the blue arrow marks the peak for a doped TBLG.}}
		\label{fig:doping2P}
	\end{figure}
	
	The parameterized form of the doping and twist angle-dependent Hartree potential energy provided by Goodwin \emph{et al.}~\cite{Goodwin2020} is discussed in Sec.~\ref{sec:IntrotoTBG}. We study the effect of the doping level on $ \alpha_1 $, and the results for electron and hole doping are presented in Figs.~\ref{fig:doping}~(a) and (b), respectively, for an arbitrary twist angle, $ \theta \sim 1.54^{\circ} $. The inset of the figure shows the behavior at low energy values. For both cases, we find that the behavior of $\alpha_1$ \emph{vs.} $E_l$ remains essentially unaltered but with a minor red shift of its resonant peaks. The red arrow indicates the energy value of $ \sim 0.185~\si{\electronvolt}$ for undoped TBLG. This value changes to $ \sim 0.16~{\si{\electronvolt}}$ and $\sim 0.2~{\si{\electronvolt}}$ (indicated by the blue arrows in FIG.~\ref{fig:doping}) for higher electron and hole filling factors, respectively. Therefore, this spectroscopic method may provide an experimental means to establish a relatively undoped sample for an arbitrary twist angle.
	
	\subsection{The two-photon absorption coefficient as a function of the incident laser energy}
	\begin{figure*}
		\centering
		\includegraphics[scale=0.22]{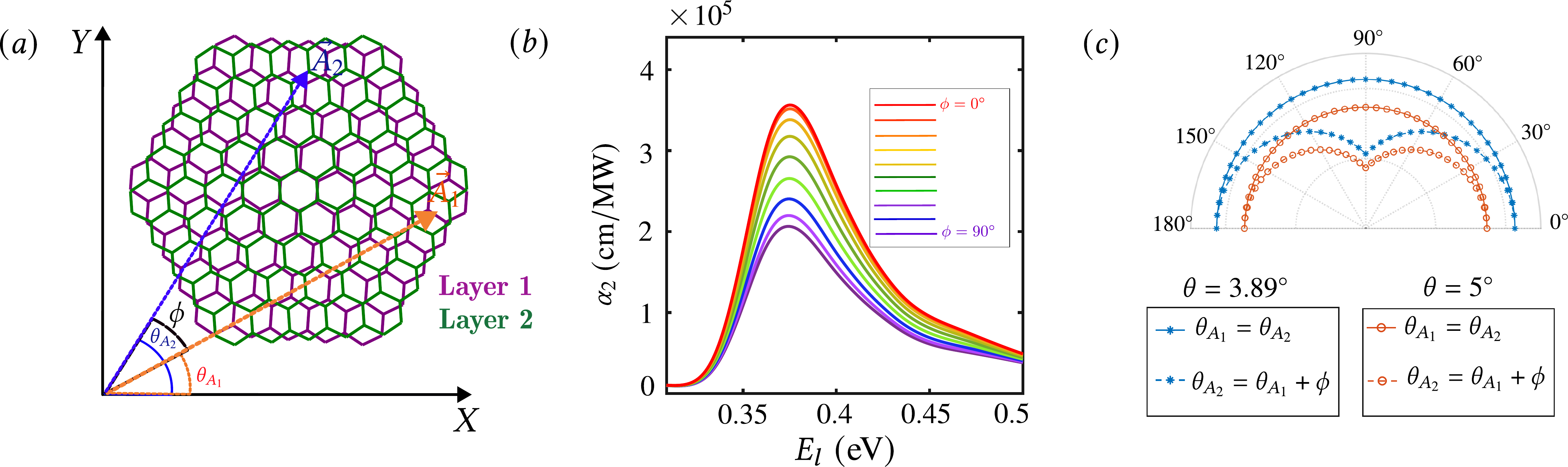}
		\caption{(a) \tcm{The magenta and green honeycomb lattices represent the real-space structures of layers 1 and 2, respectively, with a relative twist between them. The vector potentials of the two incident photons are denoted by $\vec{A}_1$ and $\vec{A}_2$, respectively, such that the phase difference between them is $\phi$. (b) The variation of the two-photon absorption coefficient $\alpha_2 $ as a function of the incident excitation energy $ E_l $ for $\theta = 3.89^{\circ} $. The polarization directions of the two incident photons, $ \theta_{A_1}$ and $ \theta_{A_2}$ are varied independently such that $ \theta_{A_2} = \theta_{A_1} + \phi $ and $\phi$ is increased from $ 0^{\circ} $ to $ 90^{\circ} $ in steps of $ 10^{\circ} $. (c) The area under the curve for $ \alpha_2 $ as a function of incident energy $ E_l $ for $ \theta = 3.89^{\circ} $ and $ 5^{\circ} $. The polar axis marks the phase difference $ \phi $ while the radial axis represents area under the curve for $ \alpha_2 $ \emph{vs} $ E_l $.}}
		\label{fig:cascadebeta}
	\end{figure*}
	By considering both the possible processes of the two-photon absorption presented in FIG.~\ref{fig:feynman}, we present the behavior of $ \alpha_2 $ (discussed in Sec.~\ref{sec:theory}) $vs. $ $E_l$ in FIG.~\ref{fig:2PA} for various $ \theta $. Our results show that a resonant peak characterizes $ \alpha_2 $ at twice the wavelength or half the energy the resonance occurs for the \tcm{one-photon} case. $ \emph{For e.g.} $, \tcm{for $ \theta \sim 5^{\circ} $, FIG.~\ref{fig:2PA} shows that the resonant feature for $ \alpha_2 $ occurs at $ E_o \sim 0.5~\si{\electronvolt} $, which is exactly half the value obtained for $ \alpha_1 $. Similarly, for $ \theta \sim 2.5^{\circ} $, $ \alpha_2 $ has a resonant peak at $\sim0.21~\si{\electronvolt}$, while for $ \alpha_1 $ it appears at $\sim 0.4~\si{\electronvolt}$. The insets of Fig.~\ref{fig:2PA} show the resonant feature's details. As the twist angle decreases, the magnitude of the resonant peak significantly increases. For instance, when $ \theta $ is reduced from $ 5^{\circ} $ to $ 1.8^{\circ} $, the resonant peak is enhanced by approximately two orders of magnitude.}
	
	Similar to the one-photon absorption process, to identify the location of wavevectors that give the significant contribution to the resonant peak in $ \alpha_2 $ \emph{vs} $ E_l $, we present the contour plot of the transition probability corresponding to the two-photon absorption as a function of $ k_x $ and $ k_y $ in the entire first MBZ in FIGS.~\ref{fig:2PAresonaceXX},~\ref{fig:2PAresonaceYY},~\ref{fig:2PAresonaceXY} for different polarization directions and mark the location of the dominant wavevectors in the bright color. Their location stabilizes for higher twist angles like $\theta \gtrsim 3.89^{\circ}$ and starts confining for smaller twist angles.
	
	\tcm{In FIG.~\ref{fig:TPAslgblgtbg}~(a), we present a clear comparison of the TPA coefficient behavior, numerically calculated for SLG, two uncoupled SLG sheets, and BLG by employing the continuum model, as reported in~\cite{Yang2011}. SLG's TPA coefficient $ \alpha_2 $ exhibits a $\omega^{-4}$ dependence by accounting for all possible transitions. The behavior for two uncoupled graphene sheets follows the same dependence but has a coefficient twice that of a single SLG sheet. The presence of two additional bands differentiates the behavior of BLG from SLG. By considering the four possible transitions (from two valence bands to two conduction bands), we see that $ \alpha_2 $ for BLG shows a strong resonant absorption peak in the infrared region (around $\lambda\sim3100~\si{\nano\meter})~\text{or}~\hbar\omega\sim0.4~\si{\electronvolt}$}.
		
	\tcm{The TPA coefficient $ \alpha_2  $ of TBLG \tcm{as a ratio against} SLG and BLG, is provided in FIG.~\ref{fig:TPAslgblgtbg}~(a) and (b), respectively. For $ \theta = 3.89^{\circ} $, $ \alpha_2 $ shows a notable enhancement over SLG by a factor of $ \sim 23 $ at $ E_l\sim 0.4~\si{\electronvolt} $. However, this enhancement decreases as the twist angle increases. When comparing $ \alpha_2 $ for TBLG with BLG, the magnitude of enhancement is significantly weaker than that observed in comparison to SLG. This increase in $ \alpha_2 $ for TBLG can be attributed to the enhanced DOS.}\tcm{For comparison, we provide the values of $ \alpha_2 $ for a few 2D materials in Table~I.}

	The effect of electron and hole doping on the behavior of $ \alpha_2 $ is shown in FIG.~\ref{fig:doping2P}~(a) and (b), respectively. On doping, we find a minor red-shift of the resonant peaks in $ \alpha_2 $ $ \emph{vs} $ $ E_l $. For electron doping, the peaks shift to the left, while for the case of hole doping, the peaks transition towards the right. The red arrow marks the energy corresponding to the resonant peak for an undoped TBLG.
	
	To study the effect of the polarization direction of the two driving photons on $ \alpha_2 $, we define the polarization directions of the two photons as $ \theta_{A_1} $ and $ \theta_{A_2} $, respectively, measured relative to an unrotated graphene sheet (refer Fig.~\ref{fig:cascadebeta}~(a)). Firstly, we vary both $ \theta_{A_1} $ and $ \theta_{A_2} $ simultaneously in-phase and secondly, we keep $ \theta_{A_1} $ fixed and vary $\theta_{A_2}$ as $ \theta_{A_2} = \theta_{A_1} + \phi $, where we scan $ \phi $ from $ 0^{\circ} $ to $ 90^{\circ} $. Here, $\phi$ marks the phase difference between the polarization of the incident photons. We observe that $ \alpha_2 $ is isotropic for the former configuration. However, for the latter, our results show that when $ \phi = 90^{\circ} $, \emph{i.e.}, both the photons are cross-polarized \textit{w.r.t} each other, $ \alpha_2 $ attains a minimum value. In contrast, $ \alpha_2 $ attains a maxima for the parallel-polarized configuration, \emph{i.e.}, when $ \theta_{A_1}=\theta_{A_2} $. In FIG.~\ref{fig:cascadebeta}~(b), we represent the result for a sample twist angle $ \theta \sim 3.89^{\circ}$ where we see that as $ \phi $ is increased from $ 0^{\circ} $ to $ 90^{\circ} $, the value of $ \alpha_2 $ keeps on decreasing. This is due to  $ \emph{interference effects}$~\cite{Narula2011,RNarulathesis2011, Arora2023} mediated by the optical matrix elements appearing in Eq.~\ref{eq:transprob}. The area under the curve for $ \alpha_2$ \textit{vs.} $ E_l $ for $ \theta \sim 3.89^{\circ},5^{\circ} $ is plotted in FIG.~\ref{fig:cascadebeta}~(c) and displays an isotropic nature when the polarization direction of the two photons are varied simultaneously and in phase. However, when rotated independently, the area under the curve exhibits an anisotropic behavior, giving maxima at $ \phi = 0^{\circ} $ and minima at $ \phi = 90^{\circ} $.
	
	\begin{center}
		\begin{table}
			\begin{tabular}{||c c c||} 
				\hline
				Material & $ \hbar\omega~(\si{\electronvolt}) $ &$ \alpha_2~\si{(\centi\meter/\mega\watt)} $\\  [0.5ex] 
				\hline\hline
				\hline
				\ch{WS_2}~\cite{Zheng}&1.54&$3.7\times10^2$\\
				\hline
				\ch{MoS_2}~\cite{Li2015}&1.2&$7.62$\\
				\hline
				\ch{MoSe_2}~\cite{Xie:19}&1.2&1.8\\
				\hline
				\ch{MoS_2}~\cite{app5041440}&2.3&$5\times10^{-4}$\\
				\hline
				\ch{WSe_2}~\cite{Liu_2018}&1.16&7.29\\[1ex]
				\hline		
			\end{tabular}
			\caption{Values of the two-photon absorption coefficients for a selection of \tcm{2D} materials at different laser energies.}
			\label{table:alpha2}
		\end{table}
	\end{center}
	\section{Conclusion} \label{sec:conc}
	In summary, our research delved into the absorption properties of twisted bilayer graphene (TBLG), specifically focusing on single- and two-photon absorption processes involving different stacking configurations and polarization orientations. We employed the effective continuum model for our analysis. Our findings revealed distinctive features for the absorption coefficient $\alpha_1$, which exhibited a series of absorption peaks corresponding to the vHs within the band structure. For TPA in TBLG, we found the presence of a resonance peak for $\alpha_2$ at nominally half the energy of the single-photon case. \tcm{At low twist angles, the magnitude of the resonant peak of TBLG responds more strongly than BLG due to an increased density of states (DOS). Nevertheless, its magnitude approaches that of two decoupled single-layer graphene (SLG) sheets as the twist angle \tcm{is increased}, \emph{e.g.}, at a twist angle of $ 2.5^{\circ} $,  the magnitude of the resonant peak for is enhanced by a factor of $ \sim 1.6 $ compared to BLG. However, for $ \theta \sim 6.5^{\circ} $ this factor reduces to $ \sim 1.05 $. A further increase in the twist angle leads to a gradual reduction in this factor.}
   \tcm{On the other hand, the TPA coefficient $ \alpha_2 $ shows distinct peaks at double the wavelength (or half the energy) at which the resonant peak for $ \alpha_1 $ appears. \emph{E.g.}, $ \alpha_1 $  shows the resonant peak for $ \theta \sim 6.5^{\circ} $ at $1.3 ~\si{\electronvolt} $, while $ \alpha_2 $ exhibits it at $ \sim 0.65~\si{\electronvolt} $. At these energies, $\alpha_2$ exhibits an enhancement of approximately an order of magnitude greater than that of SLG for $ \theta \lesssim 8^{\circ} $. Additionally, there is a small but noteworthy increase in the magnitude of $ \alpha_2 $ when compared to BLG.} 
	
	Furthermore, as we varied the twist angle, the resonant feature associated with $\alpha_1$ and $\alpha_2$ shifted from the infrared to visible excitation energies. We also explored the impact of doping on the absorption coefficient. We noted that, as the doping level increased, the behavior of $\alpha_1$ and $\alpha_2$ concerning the excitation energy $E_l$ remained consistent, accompanied by a minor red-shift of their resonant peaks. Our findings suggest that this spectroscopic technique can serve as an experimental method for determining the relative rotation angle of TBLG and identifying samples with relatively low doping levels.
	In addition to these observations, we investigated the behavior of $\alpha_2$ under various polarization configurations. Our analysis revealed that $\alpha_2$ exhibited isotropic and anisotropic behaviors when the two photons' polarization directions were altered simultaneously or independently, respectively.
	
It would be interesting to experimentally determine the value of the absorption coefficients for TBLG with the use of quantum cascade lasers~\cite{Piccardo2018,Kang2017,Dean2014,Chang2017,Deyasi17} which are capable of emitting mid-infrared light. By varying the twist angle, the resonance in the absorption coefficients of TBLG can be tuned to different energy regimes, making it an exciting prospect for the engineering of solar panels.
	
	\section{Acknowledgment}
	D. Arora acknowledges the Department of Science and
Technology, Government of India, for supporting this work
via the INSPIRE fellowship scheme. D. Aggarwal is sup-
ported by the UGC, Government of India Fellowship.
	
	\appendix
	\section{Current density operator in twisted bilayer graphene}
	\label{sec:appendix1}
	As mentioned in the Sec.~\ref{sec:theory} of the main text, an alternative approach to evaluating the absorption coefficients is by first calculating the macroscopic current operator $\textbf{j}$~\cite{Haug2009} which is determined by
	calculating the product of the velocity operator \textbf{v} and density matrix $ \rho $. In the tight-binding formalism~\cite{PhysRev.94.1498}, the Hamiltonian is constructed from the Bloch sums that span the different sublattices (say $ N $) constituting the lattice. In the basis set of these Bloch sums, the density matrix $ \rho $, the operator $ \textbf{j} $, is of $ N\times N $ dimensions. However, in moir\'e systems like TBLG~\cite{Deepanshu2023}, the effective continuum Hamiltonian~\cite{Bistritzer2011} acts on the Bloch states, constituting an infinite basis set. In reciprocal space, the Hamiltonian operator will thus consist of infinite matrix elements. The number of reciprocal lattice vectors is truncated to obtain a finite-size matrix so that all the bands in a specific energy window (chosen well beyond the incident laser energy) are converged to a predetermined tolerance. In this section, we evaluate the current operator for TBLG by employing the low energy effective continuum model~\cite{Bistritzer2011}.
	
	The real-space Hamiltonian of TBLG is given as
	\begin{widetext}
		\begin{equation}
			H_{\text{TBLG}}=\sum_{\ell=1,2}\int \dd[2]{\textbf{r}}\hat{\Psi}^{(\ell)\dagger}(\textbf{r})\,\Big[v_{F}\bm{\sigma}_{l}\cdot\left(-i\hbar\bm{\nabla} + e\textbf{A}(\bm{r})\right) + V_{H}(\bm{r})\Big]\,\hat{\Psi}^{(\ell)}(\bm{r})
			+ \int \dd[2]{\textbf{r}} \left[\hat{\Psi}^{(1)\dagger}(\textbf{r}) T'^{\dagger}(\textbf{r})\,\hat{\Psi}^{(2)}(\textbf{r}) + \text{h.c.} \right]
			\label{mat:rham1}
		\end{equation}
	\end{widetext}
	The description of each term is provided in Sec.~\ref{sec:IntrotoTBG} of the main text.
	To calculate the current density operator in TBLG, we first separate the unperturbed Hamiltonian and the perturbation due to light-matter interaction. In the second quantized form, the unperturbed $ H^0_{\text{TBLG}} $ and perturbed $ H^1_{\text{TBLG}} $ expressions of the Hamiltonian in Eq.~\ref{mat:rham1} are
	\begin{widetext}
		\begin{align}
			H^{0}_{\text{TBLG}} & =\sum_{\ell=1,2}\int \dd[2]{\textbf{r}}\hat{\Psi}^{(\ell)\dagger}(\textbf{r})\,\Big[-i\hbar v_{F}\bm{\sigma}^{(\ell)}\cdot\bm{\nabla} + V_{H}(\bm{r})\Big]\,\hat{\Psi}^{(\ell)}(\bm{r})
			+ \int \dd[2]{\textbf{r}} \left[\hat{\Psi}^{(1)\dagger}(\textbf{r}) T'^{\dagger}(\textbf{r})\,\hat{\Psi}^{(2)}(\textbf{r}) + \text{h.c.} \right] \\
			H_{1}^{\text{TBLG}} & = \sum_{\ell=1,2}\int \dd[2]{\textbf{r}}\hat{\Psi}^{(\ell)\dagger}(\textbf{r})\,\Big[e v_{F}\bm{\sigma}^{(\ell)}\cdot\textbf{A}(\bm{r}) \Big]\,\hat{\Psi}^{(\ell)}(\bm{r})
			\label{mat:rham3}
		\end{align}
		where the Pauli-matrices $\bm{\sigma}^{(\ell)}$ account for the rotation of the layers. The perturbed Hamiltonian $H^{1}_{\text{TBLG}}$ can equivalently be represented using the complete set of the Bloch states of the unperturbed Hamiltonian $H^{0}_{\text{TBLG}}$ by writing the field operators as
		\begin{equation}
			\hat{\Psi}^{(\ell)}(\bm{r}) = \sum_{n,\bm{k}} \Psi^{(\ell)}_{n\textbf{k}}(\textbf{r}) \hat{c}^{(\ell)}_{n\textbf{k}}\qq{and}			
			\hat{\Psi}^{(\ell)\dagger}(\bm{r})= \sum_{m,\textbf{k}'} \Psi^{(\ell)\dagger}_{m\textbf{k}'}(\textbf{r}) \hat{c}^{(\ell)\dagger}_{m\textbf{k}'}
		\end{equation}
		where $\Psi^{(\ell)}_{n\bm{k}}(\bm{r})$ is the Bloch function contributed by layer-$\ell$ corresponding to the band index $n$ and the Bloch wave-vector $\textbf{k}$, and $\hat{c}^{(\ell)\dagger}_{n\textbf{k}}~ (\hat{c}^{(\ell)}_{n\textbf{k}})$ are the creation (annihilation) operator corresponding to layer $ \ell $, band index $ n $ and momentum \textbf{k}. In general, the representation of the perturbation $H^{1}_{\text{TBLG}}$ is not diagonal and is written as
		\begin{equation}
			H^{1}_{\text{TBLG}} =  e v_{F} \sum_{\ell=1,2}\sum_{n,\textbf{k}}\sum_{m,\textbf{k}'}     \hat{c}^{(\ell)\dagger}_{m\bm{k}'} \int\dd[2]{\bm{r}} \Psi^{(\ell)\dagger}_{m\bm{k}'}\left[ \bm{\sigma}^{(\ell)}\cdot\textbf{A}\right] \Psi^{(\ell)\dagger}_{m\textbf{k}'}(\textbf{r}) \hat{c}^{(\ell)}_{n\textbf{k}}
		\end{equation}
	\end{widetext}
	Restricting the reciprocal lattice vector $ \textbf{G} = 0, \textbf{G}_1, -\textbf{G}_1-\textbf{G}_2$, we get the $ 8\times8 $ matrix representation of perturbed Hamiltonian as
	\begin{small}
		\begin{eqnarray}
			\frac{H_1^{TBLG}}{ev_F}=
			\begin{bmatrix}
				\bm{\sigma}_{\theta/2}\cdot\textbf{A}&0&0&0\\
				0&\bm{\sigma}_{-\theta/2}\cdot\textbf{A}&0&0\\
				0&0&\bm{\sigma}_{-\theta/2}\cdot\textbf{A}&0\\
				0&0&0&\bm{\sigma}_{-\theta/2}\cdot\textbf{A}
			\end{bmatrix}
			\label{mat:pertham}
		\end{eqnarray}
	\end{small}	
	The current
	density operator~\cite{Wang2016,Landau1981Quantum,Sipe} in the second-quantized formalism is given as
	\begin{eqnarray}
		\textbf{j}(\textbf{r}) =-e
		\begin{bmatrix}
			\Psi^{\dagger}_{1}(\textbf{r})&\Psi^{\dagger}_{2}(\textbf{r})
		\end{bmatrix}
		\textbf{v}
		\begin{bmatrix}
			\Psi_{1}(\textbf{r})\\ \Psi_{2}(\textbf{r})
		\end{bmatrix}
	\end{eqnarray}
	where, $\textbf{v}$ is the velocity operator defined as $ =\frac{i}{\hbar}\comm{H}{r} $~\cite{Parker2019} is given by
	\begin{eqnarray}
		\small
		\frac{\textbf{v}}{(i/\hbar)}=
		\begin{bmatrix}
			\comm{	h_{1}\left(-\theta/2\right)+V_H(\textbf{r})}{\textbf{r}}&\comm{T'^{\dagger}(\textbf{r})}{\textbf{r}}\\
			\comm{T'(\textbf{r})}{\textbf{r}}&\comm{	h_{2}\left(\theta/2\right)+V_H(\textbf{r})}{\textbf{r}}
		\end{bmatrix}
	\end{eqnarray}
	Solving each term in the above matrix, we get
	\begin{eqnarray*}
		\comm{	h_{1}\left(-\theta/2\right)+V_H(\textbf{r})}{\textbf{r}} &=& \comm{	h_{1}\left(-\theta/2\right)}{\textbf{r}} +\comm{	V_H(\textbf{r})}{\textbf{r}} \\ \nonumber
		&=&\comm{h_{1}\left(-\theta/2\right)}{\textbf{r}}\\
		\noindent\text{and}\\
		\comm{	h_{1}\left(-\theta/2\right)+V_H(\textbf{r})}{\textbf{r}} &=& \comm{	h_{1}\left(-\theta/2\right)}{\textbf{r}}
	\end{eqnarray*}	
	while $
	\comm{T'(\textbf{r})}{\textbf{r}} = \comm{T'^{\dagger}(\textbf{r})}{\textbf{r}} = 0
	$.\\
	
	Therefore, the velocity operator takes the form
	\begin{eqnarray}
		\textbf{v}=\frac{i}{\hbar}
		\begin{bmatrix}
			\comm{	h_{1}\left(-\theta/2\right)}{\textbf{r}}&0\\
			0&\comm{	h_{2}\left(\theta/2\right)}{\textbf{r}}
		\end{bmatrix}
		\label{eq:curr}
	\end{eqnarray}
	Using the commutation between the rotated Dirac Hamiltonian and the position operator as
	\begin{eqnarray}
		\frac{i}{\hbar}\comm{	h_{1}\left(-\theta/2\right)}{\textbf{r}} = v_F\bm{\sigma}_{\theta/2}\\ \nonumber
		\frac{i}{\hbar}\comm{	h_{2}\left(-\theta/2\right)}{\textbf{r}} = v_F\bm{\sigma}_{-\theta/2}
	\end{eqnarray}
	we can rewrite Eq.~\ref{eq:curr} as
	\begin{eqnarray}
		\textbf{v}=v_F
		\begin{bmatrix}
			\bm{\sigma}_{\theta/2}&0\\
			0&\bm{\sigma}_{-\theta/2}
		\end{bmatrix}
	\end{eqnarray}
	Defining $ \ket{n}=\ket{\textbf{k},n} $, for the Hamiltonian $ H=H_0 + H_1$, leads to the von Neumann equation for the density matrix ~\cite{BOYD2008277,Wang2016}
	\begin{equation}
		\small
		i\hbar\frac{\partial}{\partial t}\rho_{mn}=\left[E_m(\textbf{k})-E_n(\textbf{k})\right])\rho_{mn}+\sum_l\left[(H_1)_{ml}\rho_{ln}-\rho_{ml}(H_1)_{ln}\right]
		\label{Eq:vn}
	\end{equation}
	The macroscopic current~\cite{Gao2021,Wang2016,Haug2009} is computed as
	\begin{equation}
		\textbf{j}(t)=-e	\sum_{mn}\textbf{v}_{mn}\rho_{mn}(\textbf{k},t)
	\end{equation}
	where $ \textbf{v}_{mn} = \mel{m}{\textbf{v}}{n} $ are the matrix elements of the velocity operator.
	The current operator up to linear and second order in the electric field is given as~\cite{Wang2016,Gao2021}
	\begin{eqnarray}
		\textbf{j}^{1}(t)&=&-e	\sum_{mn}\textbf{v}_{mn}\rho^{(1)}_{mn}(\textbf{k},t)~\text{and}\\ \nonumber
		\textbf{j}^{2}(t)&=&-e	\sum_{mn}\textbf{v}_{mn}\rho^{(2)}_{mn}(\textbf{k},t)
	\end{eqnarray}
	In the above expression, $ \rho^{(1)}_{mn}(\textbf{k},t) $ is the solution of the density matrix equation in the linear
	approximation with respect to the field and is given as~\cite{Gao2021}
	\begin{equation}
		\rho^{(1)}_{nm}(\textbf{k},t) = e\frac{(f_n^{T}(\textbf{k},\mu)-f_m^{T}(\textbf{k},\mu))(\textbf{v}\cdot\textbf{A})_{nm}}{E_n(\textbf{k})-E_m(\textbf{k})+\hbar\omega}e^\eta t
		\label{eq:rho1}
	\end{equation}
	where, $V_{nm}=(\textbf{v}\cdot\textbf{A})_{nm}  $ are the matrix elements of $ H_1 $ given in Eq.~\ref{mat:pertham} in the Bloch basis and $\eta\rightarrow0^+$ denotes the perturbation is turned on adiabatically at $t\rightarrow-\infty$. The zeroth order contribution to the density matrix
	$\rho^{(0)}_{nm} = \delta_{mn}f^T_n(\textbf{k},\mu)$ describes the occupation of the electrons in the ground state before the application of an external electric field to the system, which in the Bloch basis is the Fermi occupation functions denoted by $ f^T(\textbf{k},\mu) $. The second order solution to Eq.~\ref{Eq:vn} \textit{i.e.} the term quadratic with respect to the field can be written as~\cite{Wang2016}
	\begin{widetext}
		\begin{eqnarray}
			\rho^{(2)}_{nm}(\textbf{k},t) &=& e^2\sum_{l,\omega_1,\omega_2}\frac{(f^T_m(\textbf{k},\mu)-f^T_l(\textbf{k},\mu))(\textbf{v}\cdot\textbf{A})_{nl}(\textbf{v}\cdot\textbf{A})_{lm}}{(E_m(\textbf{k})-E_l(\textbf{k})+\hbar\omega_1-i\hbar\eta)(E_n(\textbf{k})-E_m(\textbf{k})+\hbar(\omega_1+\omega_2)-2i\hbar\eta)}\\
			&+&\frac{(f^T_n(\textbf{k},\mu)-f^T_l(\textbf{k},\mu))(\textbf{v}\cdot\textbf{A})_{nl}(\textbf{v}\cdot\textbf{A})_{lm}}{(E_n(\textbf{k})-E_l(\textbf{k})+\hbar\omega_1-i\hbar\eta)(E_n(\textbf{k})-E_m(\textbf{k})+\hbar(\omega_1+\omega_2)-2i\hbar\eta)}e^{2\eta t}\nonumber
			\label{eq:rho2}
		\end{eqnarray}
	\end{widetext}
	\begin{figure}
		\includegraphics[scale=0.5]{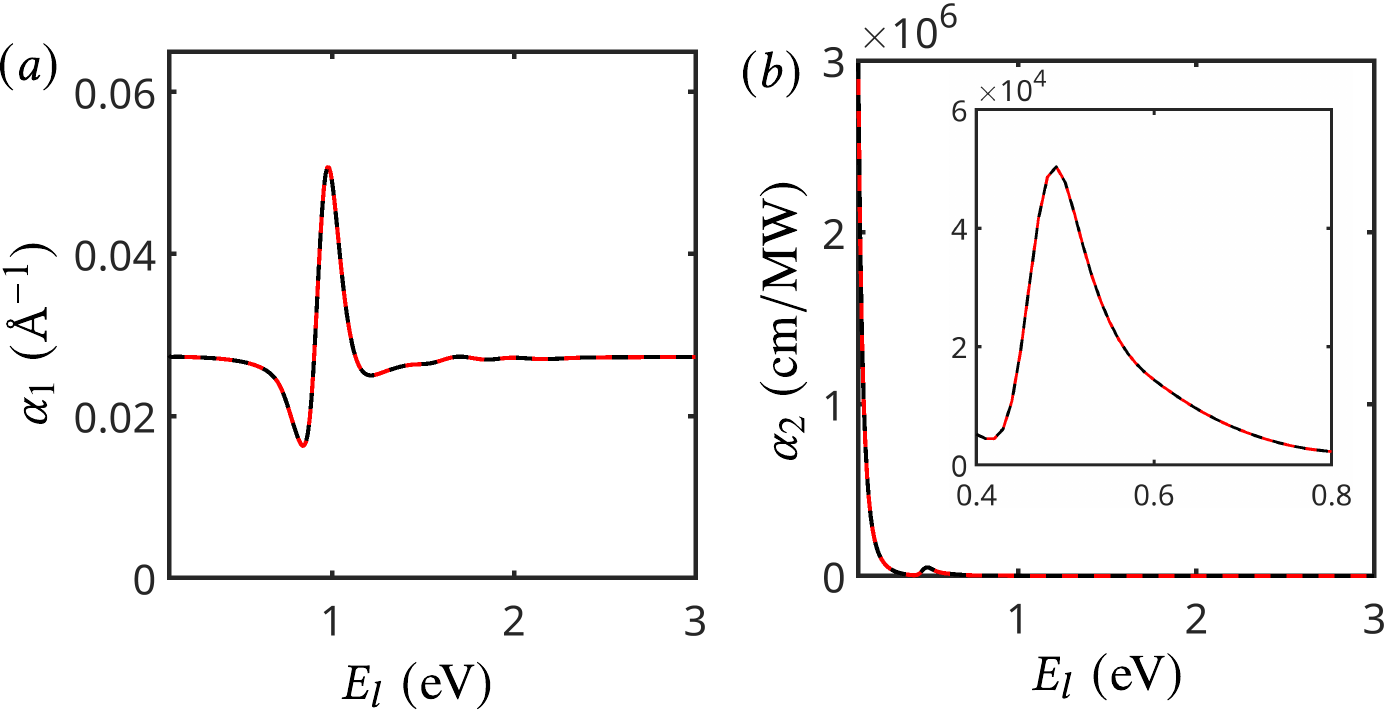}
		\caption{(a) One- and (b) two-photon absorption coefficients as a function of incident excitation energy for $ \theta\sim5^{\circ} $, respectively. The red solid and black dashed lines represent the absorption coefficients evaluated by Eq.~\ref{eq:alpha} in the main text and macroscopic current operator, respectively. \tcm{The inset of (b) shows the \tcm{details} for lower $ \hbar\omega $.}}
		\label{fig:alpha}
	\end{figure} 
	The absorption coefficient for incident light with frequency $ \omega $ is given as~\cite{phdthesisWeber,Graphene2013,Malic2006}
	$\alpha(\omega)\approx\frac{1}{\epsilon_o}\Im\left[j(\omega)/A(\omega)\right]$. In Fig.~\ref{fig:alpha}, we have provided the plots for $ \alpha_{i=1,2} $ and compared them with our results obtained from the methodology described in Sec.~\ref{sec:theory}. We see that both the approaches give identical behavior for $ \alpha_{i=1,2} $. For illustration, we provide the expression for the absorption coefficient using both formalisms for graphene in the linear band approximation.
	The transition probability from an initial valence band to a final conduction band for a single photon absorption is given as:
	\begin{align}
		w_{i\rightarrow f}&=\frac{2\pi}{\hbar}\int_{\textbf{k}}\abs{\mel{\Psi_{\textbf{k},f}}{H_P}{\Psi_{\textbf{k},i}}}^2\delta(E_f(\textbf{k})-E_i(\textbf{k})-\hbar\omega)\\ \nonumber
		&\times\left[f^T_i(\textbf{k},\mu)\left\{1-f^T_f(\textbf{k},\mu)\right\}\right]\dd^2\textbf{k}\nonumber
	\end{align}
	where $ f^T_i(\textbf{k},\mu) $ is the probability of the initial state $ i $ being filled and $ 1-f^T_f(\textbf{k},\mu) $ is the probability of the final state $ f $ being empty. At $ T = 0~K$, the term $ f^T_i(\textbf{k},\mu)\left[1-f^T_f(\textbf{k},\mu)\right] = 1 $. For finite temperature, we have the explicit form of the Fermi functions given as
	\begin{equation}
		f^T_i(\textbf{k},\mu)=\frac{1}{\exp(\frac{E_i(\textbf{k})-\mu}{k_BT})+1}
	\end{equation}
	where $ T $ is the temperature and $ k_B $ is the Boltzmann constant. 	Considering the linear band dispersions for graphene as $
	E_s(\textbf{k}) = s\hbar v_Fk\qq{and}E_v(\textbf{k})
	$, where $ s $ is the band index with value $ s= +1(-1) $ for conduction and valence band, respectively, we can rewrite $w_{i\rightarrow f}$ as
	\begin{eqnarray*}
		w_{i\rightarrow f}&=&\frac{1}{2\pi\hbar}\int_{\textbf{k}}\abs{\mel{\Psi_{\textbf{k},c}}{H_P}{\Psi_{\textbf{k},v}}}^2\delta(2\hbar v_Fk-\hbar\omega)\\
		&\times&\frac{\exp(\frac{\hbar v_Fk-\mu}{k_BT})}{\left[\exp(\frac{-\hbar v_Fk-\mu}{k_BT})+1\right]\left[\exp(\frac{\hbar v_Fk-\mu}{k_BT})+1\right]}\dd^2\textbf{k}
	\end{eqnarray*}
	On solving the integration over $ k $, we get the first-order transition probability
	\begin{equation}
		w_{i\rightarrow f}=\frac{A^2e^2\omega}{8\hbar^2}\frac{\exp(\frac{\frac{\hbar\omega}{2}-\mu}{k_BT})}{\left[\exp(\frac{\frac{-\hbar\omega}{2}-\mu}{k_BT})+1\right]\left[\exp(\frac{\frac{\hbar\omega}{2}-\mu}{k_BT})+1\right]}
	\end{equation}
	The photon absorption coefficient is given as
	$	\alpha_1 =  \frac{2\hbar\omega w_{i\rightarrow f}}{Id}$ which gives
	\begin{equation}
		\alpha_1 = \frac{e^2}{2\hbar c\epsilon_o~d}\frac{\exp(\frac{\frac{\hbar\omega}{2}-\mu}{k_BT})}{\left[\exp(\frac{\frac{-\hbar\omega}{2}-\mu}{k_BT})+1\right]\left[\exp(\frac{\frac{\hbar\omega}{2}-\mu}{k_BT})+1\right]}
		\label{eq:our}
	\end{equation}
	To find the expression of $ \alpha_1 $ from the macroscopic current operator, the matrix elements of the first-order density operator are given as~\cite{Wang2016}
	\begin{eqnarray}
		\rho^{(1)}_{s's}(\omega) =\frac{ ev_f\left[\bm{\sigma}\cdot\textbf{A}\right]_{s's}\left(\rho_{ss}-\rho{s's'}\right)}{\hbar\omega-(E_{s'}-E_s)}
	\end{eqnarray}
	and the current operator, as defined previously, is given as 
	\begin{eqnarray}
		\textbf{j}_{ss'}(\omega) &=&e\sum_{\textbf{k}} \sum_{s's}\textbf{v}_{s's}\rho_{ss'}(\omega)
	\end{eqnarray}
	where $ v =v_F\bm{\sigma}$ is the velocity operator for graphene. Solving the above expression gives the absorption coefficient as
	\begin{eqnarray}
		\alpha_1(\omega) 
		= \frac{e^2}{2\hbar c\epsilon_o}\frac{\exp(\frac{\frac{\hbar\omega}{2}-\mu}{k_BT})-\exp(\frac{\frac{-\hbar\omega}{2}-\mu}{k_BT})}{\left[\exp(\frac{\frac{-\hbar\omega}{2}-\mu}{k_BT})+1\right]\left[\exp(\frac{\frac{\hbar\omega}{2}-\mu}{k_BT})+1\right]}
		\label{eq:mele}
	\end{eqnarray}
	For a finite temperature and $ \mu $, Eq.~\ref{eq:mele} reduces to Eq.~\ref{eq:our} for $ \omega >\omega_o $.
	
	Rewriting Eq.~\ref{eq:rho2} for the DC response, we have
	\begin{widetext}
		\begin{eqnarray}
			\rho^{(2)}_{nm}(\textbf{k},t) &=& e^2\sum_{l,\omega_1,\omega_2}\frac{(f^T_m(\textbf{k},\mu)-f^T_l(\textbf{k},\mu))(\textbf{v}\cdot\textbf{A})_{nl}(\textbf{v}\cdot\textbf{A})_{lm}}{(E_m(\textbf{k})-E_l(\textbf{k})+\hbar\omega_1-i\hbar\eta)(E_n(\textbf{k})-E_m(\textbf{k})-2i\hbar\eta)}\\
			&+&\frac{(f^T_n(\textbf{k},\mu)-f^T_l(\textbf{k},\mu))(\textbf{v}\cdot\textbf{A})_{nl}(\textbf{v}\cdot\textbf{A})_{lm}}{(E_n(\textbf{k})-E_l(\textbf{k})+\hbar\omega_1-i\hbar\eta)(E_n(\textbf{k})-E_m(\textbf{k})-2i\hbar\eta)} \nonumber e^{2\eta t}
		\end{eqnarray}
	\end{widetext}
	
	The total current operator can be written as a summation of diagonal and off-diagonal terms, \textit{i.e.}, $
	\textbf{j}  = \textbf{j}^{\text{dia}} + \textbf{j}^{\text{off}}$ 
	where, $\textbf{j}^{\text{dia}}$ and $\textbf{j}^{\text{off}}$ are defined as
	\begin{eqnarray}
		\textbf{j}^{\text{dia}} = \frac{e}{V}\sum_{n,\textbf{k}}\textbf{v}_{nn}(\textbf{k})\rho^{(2)}_{nn}(\textbf{k},t)\\
		\textbf{j}^{\text{off}} = \frac{e}{V}\sum_{n,m,n\neq m,\textbf{k}}\textbf{v}_{mn}(\textbf{k})\rho^{(2)}_{nm}(\textbf{k},t)
	\end{eqnarray}
	To calculate the diagonal contribution to the current operator, we replace the term $ \lim_{\eta\rightarrow0}\frac{-1}{2i\hbar\eta}\frac{1}{E_n(\textbf{k})-E_m(\textbf{k})+\hbar\omega-i\hbar\eta} $ by
	\begin{eqnarray}
		&\lim&_{\eta\rightarrow0}-\frac{\pi}{2\hbar\eta}\delta(E_n(\textbf{k})-E_m(\textbf{k})+\hbar\omega)\\ \nonumber&-&\frac{1}{2i\hbar\eta}\mathcal{P}\frac{1}{E_n(\textbf{k})-E_m(\textbf{k})+\hbar\omega}-\frac{1}{2}\mathcal{P}\frac{1}{(E_n(\textbf{k})-E_m(\textbf{k})+\hbar\omega)^2}
	\end{eqnarray}
	Using the above expression $ j^{\text{dia}} $ can be broken into three parts \textit{i.e.}, $ \textbf{j}^{\text{dia}} = \textbf{j}^{\text{dia1}} + \textbf{j}^{\text{dia2}}+\textbf{j}^{\text{dia3}} $.
	By using the above identity, we obtain the three diagonal terms as
	\begin{widetext}
		\begin{eqnarray*}
			\textbf{j}^{\text{dia1}} &=&- \frac{e^3\pi}{2V\hbar\eta}\sum_{n,m,i,j,\omega,\textbf{k}}(f^T_n(\textbf{k},\mu)-f^T_m(\textbf{k},\mu))\textbf{v}_{nn}((\textbf{k})-\textbf{v}_{mm}(\textbf{k}))(\textbf{v}\cdot\textbf{A})_{nl}(\textbf{v}\cdot\textbf{A})_{lm}\delta(E_n(\textbf{k})-E_m(\textbf{k})+\hbar\omega)\\ \nonumber
			\textbf{j}^{\text{dia2}} &=& -\frac{e^3}{V2i\hbar\eta}\sum_{n,m,\textbf{k}}(f^T_n(\textbf{k},\mu)-f^T_m(\textbf{k},\mu))(\textbf{v}_{nn}(\textbf{k})-\textbf{v}_{mm}(\textbf{k}))(\textbf{v}\cdot\textbf{A})_{nl}(\textbf{v}\cdot\textbf{A})_{lm}\frac{1}{E_n(\textbf{k})-E_m(\textbf{k})+\hbar\omega}\\ \nonumber	
			\textbf{j}^{\text{dia3}} &=&\frac{e^3}{V}\sum_{n,m,i,j,\omega,\textbf{k}}(f^T_n(\textbf{k},\mu)-f^T_m(\textbf{k},\mu))
			\times\mathcal{P}\frac{\textbf{v}_{nn}(\textbf{k})-\textbf{v}_{mm}(\textbf{k})}{2(E_n(\textbf{k})-E_m(\textbf{k})+\hbar\omega)^2}(\textbf{v}\cdot\textbf{A})_{nm}(\textbf{v}\cdot\textbf{A})_{mn}
			\label{eq:jdia3}
		\end{eqnarray*} 
	\end{widetext}
	For $ \textbf{j}^{\text{dia3}}$, the principal part in the above expression can be replaced by
	$
	- \bm{\nabla}_{\textbf{k}}\mathcal{P}\frac{1}{2\hbar}\frac{1}{(E_n(\textbf{k})-E_m(\textbf{k})+\hbar\omega)}
	$
	which reduces the term $ \textbf{j}^{\text{dia3}} $ as
	\begin{widetext}
		\begin{eqnarray}
			\textbf{j}^{\text{dia3}}= \frac{e^3}{V2\hbar}\sum_{n,m,i,j,\omega,\textbf{k}}(f^T_n(\textbf{k},\mu)-f^T_m(\textbf{k},\mu))
			\times\bm{\nabla}_{\textbf{k}}\frac{1}{E_n(\textbf{k})-E_m(\textbf{k})+\hbar\omega}(\textbf{v}\cdot\textbf{A})_{nm}(\textbf{v}\cdot\textbf{A})_{mn}
			\label{eq:dia3}
		\end{eqnarray}
	\end{widetext}		
	\begin{widetext}
		The off-diagonal contribution to the current is given as
		\begin{align}
			\small
			\textbf{j}^{\text{off}}  &= \frac{e}{V}\sum_{n,m,n\neq m} \textbf{v}_{mn}(\textbf{k})\rho^{(2)}_{nm}(\textbf{k},\omega)\\ \nonumber
			&= \frac{e^3i}{V\hbar}\sum_{i,j,m,n,\textbf{k},\omega}\frac{(f^T_m(\textbf{k},\mu)-f^T_l(\textbf{k},\mu))(\textbf{v}\cdot\textbf{A})_{nl}(\textbf{v}\cdot\textbf{A})_{lm}}{(E_m(\textbf{k})-E_l(\textbf{k})+\hbar\omega_1-i\hbar\eta)}\textbf{r}_{mn}(\textbf{k}) + \frac{(f^T_n(\textbf{k},\mu)-f^T_l(\textbf{k},\mu))v^j_{nl}(\textbf{k})(\textbf{v}\cdot\textbf{A})_{nm}(\textbf{v}\cdot\textbf{A})_{mn}}{(E_n(\textbf{k})-E_l(\textbf{k})+\hbar\omega_1-i\hbar\eta)} \textbf{r}_{mn}(\textbf{k})\\ \nonumber
			&+ \frac{e^3i}{V\hbar}\sum_{i,j,m,n,\textbf{k}}\frac{(f^T_m(\textbf{k},\mu)-f^T_l(\textbf{k},\mu))v^i_{lm}(\textbf{k})}{(E_m(\textbf{k})-E_l(\textbf{k})+\hbar\omega_1-i\hbar\eta)}A_i(\omega)A_j(\omega)\int\dd\textbf{k}'v_{nl}^{\textbf{k}'\textbf{k}}\textbf{r}_{mn}^{\textbf{k}\textbf{k'}}+ \frac{(f^T_n(\textbf{k},\mu)-f^T_l(\textbf{k},\mu))v^i_{nl}(\textbf{k})}{(E_m(\textbf{k})-E_l(\textbf{k})+\hbar\omega_1-i\hbar\eta)}\int\dd\textbf{k}'v_{lm}^{\textbf{k}'\textbf{k}}\textbf{r}_{mn}^{\textbf{k}\textbf{k'}}
			\label{eq:joff1}
		\end{align}	
	\end{widetext}
	where in the above expression the integration limits are from $ {\textbf{k}-\epsilon}~\text{to}~ {\textbf{k}+\epsilon} $ and $ \epsilon\rightarrow 0 $.
	In the first line of Eq.~\ref{eq:joff1}, if we rotate the indices $ l\rightarrow n~, m \rightarrow l~\text{and}~n\rightarrow m $, we get the term $ \comm{v^j_{mn}\textbf{r}_{lm}}{v^j_{lm}\textbf{r}_{mn}}$ which turns out to be equal to 0 due to the commutation between $ \textbf{r} $ and $ \textbf{p} $ as $ \comm{r^i}{v^j}=i\hbar\delta_{ij} $. In Eq.~\ref{eq:joff1}, we have written the velocity $ \textbf{v} $ operator in terms of the position operator $ \textbf{r} $. To do this, we use the definition of velocity operator to calculate its matrix elements as~\cite{Juan2020}
	\begin{eqnarray*}
		v_{nm}(\textbf{k}) &=& \frac{i}{\hbar} \mel{n}{\left[H,\textbf{r}\right]}{m} 	= \frac{i}{\hbar}\left[\mel{n}{H\textbf{r}}{\chi_m(\textbf{k})} -  \mel{m}{\textbf{r}H}{m}\right]\\ \nonumber 
		&=&\frac{i}{\hbar}(E_n(\textbf{k})-E_m(\textbf{k}))\textbf{r}_{nm}(\textbf{k})
	\end{eqnarray*}
	where $ \textbf{r}_{nm}(\textbf{k}) $ is defined as $ {r}^{\textbf{k}\textbf{k}'}_{nm}= \mel{n\textbf{k}}{\textbf{r}}{m\textbf{k}'}$. These matrix elements are computed as
	\begin{align*}
		\bmr_{nm}(\bmk,\bmk') &= \mel{n\textbf{k}}{\bmr}{m\textbf{k}'}=\int\dd[d]\bmr\braket{n\textbf{k}}{\bmr}\,\bmr\,\braket{\bmr}{m\textbf{k}'}\\
		& =\int\dd[d]\bmr~u^{*}_{n\bmk}(\bmr) e^{-i\bmk\cdot\bmr}\,\bmr\,u_{m\bmk'}(\bmr) e^{i\bmk'\cdot\bmr} \\
		& =\int\dd[d]\bmr~u^{*}_{n\bmk}(\bmr) e^{-i\bmk\cdot\bmr}\,u_{m\bmk'}(\bmr)(-i\bm{\nabla}_{\bmk'} e^{i\bmk'\cdot\bmr})
	\end{align*}
	Using simply the product rule of the gradient operator, it can further be rewritten as
	\begin{widetext}
		\begin{align}
			\bmr_{nm}(\bmk,\bmk') 
			& = \int \dd[d]{\bmr} ~u^{*}_{n\bmk}(\bmr) e^{-i\bmk\cdot\bmr}\left\{(-i\bm{\nabla}_{\bmk'})\left(u_{m\bmk'}(\bmr)\,e^{i\bmk'\cdot\bmr}\right)\right\} + \int \dd[d]{\bmr} ~u^{*}_{n\bmk}(\bmr) e^{-i\bmk\cdot\bmr}\left\{ie^{i\bmk'\cdot\bmr}\,\bm{\nabla}_{\bmk'} u_{m\bmk'}(\bmr)\right\}  \\ \nonumber
			& = (-i\bm{\nabla}_{\bmk'}) \delta_{nm} (2\pi)^{d} \delta(\bmk'-\bmk) + \int\dd[d]{\bmr}~ ie^{\left(\bmk'-\bmk\right)\cdot\bmr} u^{*}_{n\bmk}(\bmr)\bm{\nabla}_{\bmk'}u_{m\bmk'}(\bmr) \\ \nonumber
			& = (-i\bm{\nabla}_{\bmk'}) \delta_{nm} (2\pi)^{d} \delta(\bmk'-\bmk) + (2\pi)^{d} \delta(\bmk'-\bmk)\textbf{r}_{nm}(\textbf{k})
		\end{align}
	\end{widetext}
	From the above expression, we see that, the first term vanishes if the band indices $ n\neq m $. For identical band indices, \textit{i.e.}, for $ n=m $, we integrate the above expression for $ \bmk = \bmk' $
	\begin{widetext}
		\begin{eqnarray}
			\int_{\textbf{k}-\epsilon}^{\textbf{k}+\epsilon}\dd^3\textbf{k}'g(\textbf{k}')\textbf{r}_{nn}^{\bmk,\bmk'}= i\bm{\nabla}_\textbf{k}g(\textbf{k})+g(\textbf{k})\textbf{R}_{nn}(\textbf{k})\qq{and}
			\int_{\textbf{k}-\epsilon}^{\textbf{k}+\epsilon}\dd^3\textbf{k}'g(\textbf{k}')\textbf{r}_{nn}^{\bmk',\bmk} = -i\bm{\nabla}_\textbf{k}g(\textbf{k})+g(\textbf{k})\textbf{R}_{nn}(\textbf{k})
			\label{eq:int1}
		\end{eqnarray}
	\end{widetext}
	where $ \textbf{r} $ has been replaced by $ \textbf{R} $.
	Using Eq.~\ref{eq:int1} in \ref{eq:joff1}, we obtain the expression for the off-diagonal contribution to the total current operator as
	\begin{widetext}
		\begin{align}
			\textbf{j}^{\text{off}}&=\frac{e^3}{2V\hbar} \sum_{n,m,i,j,\textbf{k}}\frac{f^T_m(\textbf{k},\mu)-f^T_n(\textbf{k},\mu)}{E_n(\textbf{k})-E_m(\textbf{k})+\hbar\omega}A_i(\omega)A_j(\omega)\left(v^i_{nm}(\textbf{k})\bm{\nabla}_{\textbf{k}}v^j_{mn}(\textbf{k})+(\textbf{R}_{mm}(\textbf{k})-\textbf{R}_{nn}(\textbf{k}))v^i_{nm}(\textbf{k})v^j_{mn}(\textbf{k})\right)
			\label{eq:joff2}
		\end{align}
	\end{widetext}
	Here, $ i,j $ are the Cartesian coordinates, and $v^i_{mn}$ is the velocity operator's $ i^{th} $ component. By using the following identity
	\begin{widetext}
		\begin{eqnarray}
			\small
			\lim_{\eta\rightarrow0}\frac{1}{2i\hbar\eta}\frac{1}{E_n(\textbf{k})-E_m(\textbf{k})+\hbar\omega-i\hbar\eta} =\lim_{\eta\rightarrow0}\frac{\pi}{2\hbar\eta}\delta(E_n(\textbf{k})-E_m(\textbf{k})+\hbar\omega) + \mathcal{P}\frac{1}{E_n(\textbf{k})-E_m(\textbf{k})+\hbar\omega}
			\label{eq:id}
		\end{eqnarray}
	\end{widetext}
	
\begin{figure}
		\centering
		\includegraphics[scale=0.27]{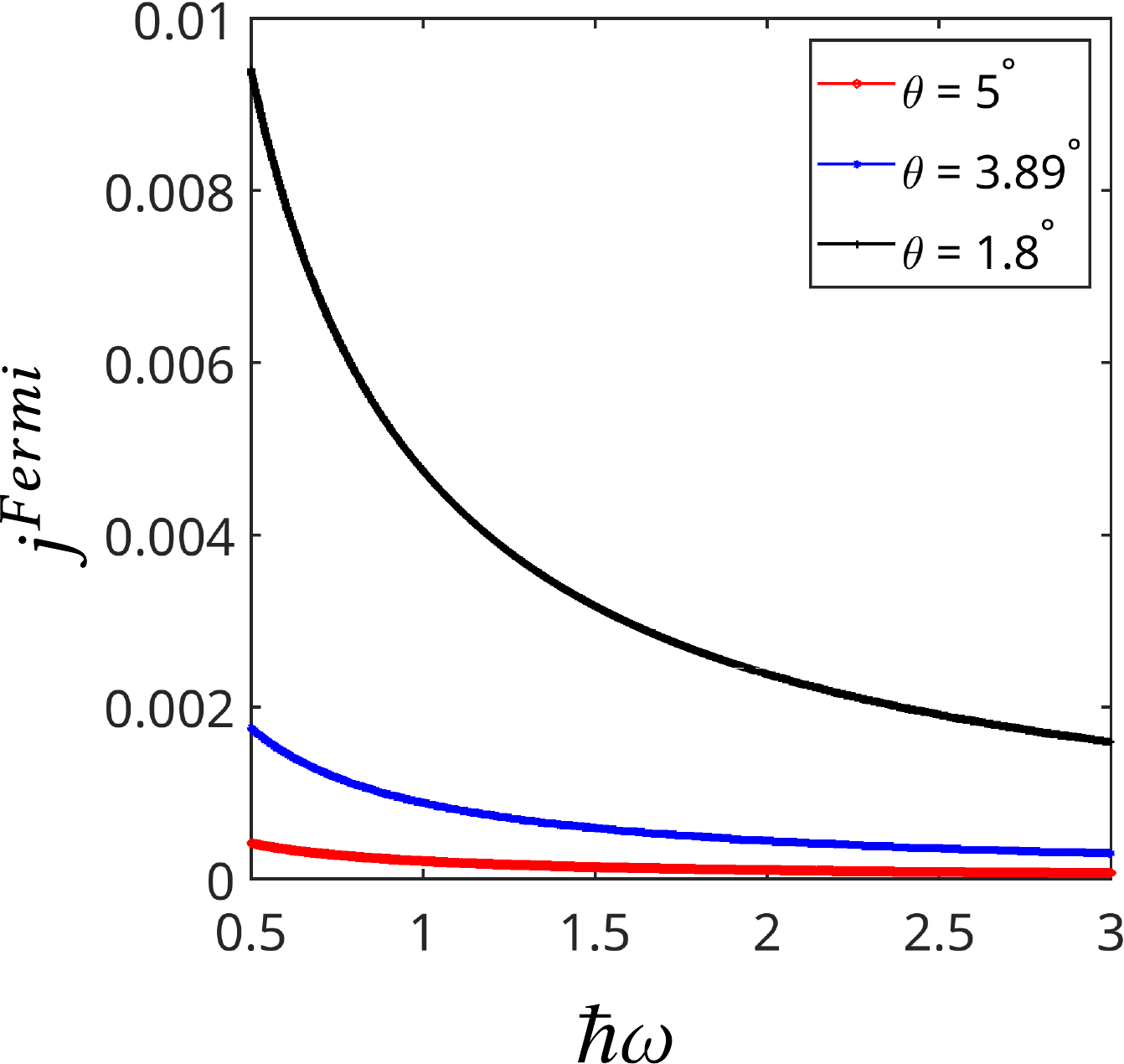}
		\caption{$j^{Fermi}$ as a function of $ \hbar\omega$ for different twist angles. The value of $ j^{\text{Fermi}} $ increases as the laser energy is reduced. At a given laser energy, lower twist angles have a higher value of $ j^{\text{Fermi}}$.}
		\label{fig:jfermi}
	\end{figure}

	We separate out the resonant $ (\textbf{j}^{\text{off1}}) $ and non-resonant contribution $ (\textbf{j}^{\text{off2}}) $ to Eq.~\ref{eq:dia3}. Using the first term on the RHS of Eq.~\ref{eq:id} in Eq.~\ref{eq:joff2} gives the resonant contribution as
	\begin{widetext}
		\begin{equation}
			\textbf{j}^{\text{off1}}=\frac{e^3\pi}{V\hbar} \sum_{n,m,i,j,\textbf{k}}(f^T_m(\textbf{k},\mu)-f^T_n(\textbf{k},\mu))A_i(\omega)A_j(\omega)
			\left(v^i_{nm}(\textbf{k})\bm{\nabla}_{\textbf{k}}v^j_{mn}(\textbf{k})+(\textbf{R}_{mm}(\textbf{k})-\textbf{R}_{nn}(\textbf{k}))v^i_{nm}v^j_{mn}\right)
			\delta(E_n(\textbf{k})-E_m(\textbf{k})+\hbar\omega)
		\end{equation}
		\text{The remaining non-resonant contribution related to the principal part is given as}
		\begin{equation}
			\textbf{j}^{\text{off2}}=\frac{e^3\pi}{V\hbar} \sum_{n,m,i,j,\textbf{k}}(f^T_m(\textbf{k},\mu)-f^T_n(\textbf{k},\mu))A_i(\omega)A_j(\omega)
			\left(v^i_{nm}(\textbf{k})\bm{\nabla}_{\textbf{k}}v^j_{mn}(\textbf{k})+(\textbf{R}_{mm}(\textbf{k})-\textbf{R}_{nn}(\textbf{k}))v^i_{nm}v^j_{mn}\right)\mathcal{P}\frac{1}{E_n(\textbf{k})-E_m(\textbf{k})+\hbar\omega}
		\end{equation}
	\end{widetext}
	On adding the non-resonant contribution to the off-diagonal current and the diagonal term of $ \textbf{j}(t) $ given in Eq.~\ref{eq:dia3} and the non-resonant part of the off-diagonal term given in the above expression in which $(\textbf{R}_{mm}(\textbf{k})-\textbf{R}_{nn}(\textbf{k}))v^i_{nm}(\textbf{k})v^j_{mn}(\textbf{k})$ vanishes, we obtain the contribution of the Fermi surface
	\begin{eqnarray}
		\textbf{j}^{\text{Fermi}}&=&\frac{e^3}{V\hbar} \sum_{n,m,\textbf{k}}\bm{\nabla}_{\textbf{k}}\left[f^T_n(\textbf{k},\mu)-f^T_m(\textbf{k},\mu)\right]\\ \nonumber
		&\times&\frac{(\textbf{v}\cdot\textbf{A})_{nm}(\textbf{v}\cdot\textbf{A})_{mn}}{E_n(\textbf{k})-E_m(\textbf{k})+\hbar\omega}
		\label{eq:jfermi}
	\end{eqnarray}
	To evaluate $ \textbf{j}^{\text{Fermi}} $ in TBLG, we need to evaluate the expression given in Eq.~\ref{eq:jfermi} for all the pairs of bands. The summation over $ \textbf{k} $ includes all the k-points in the MBZ. Below, we provide a brief explanation to demonstrate the behavior of $ \bm{\nabla}_{\textbf{k}}\left[f^T_n(\textbf{k},\mu)-f^T_m(\textbf{k},\mu)\right] $. For our purpose, we rename the indices $ n(m) $  by $ c(v) $, where $ c $ stands for conduction and $ v $ stands for valence band index and write the term $	\bm{\nabla}_{\textbf{k}}\left[f^T_c(\textbf{k},\mu)-f^T_v(\textbf{k},\mu)\right] $ as
	\begin{eqnarray*}
		\bm{\nabla}_{\textbf{k}}(F)&=&\bm{\nabla}_{\textbf{k}}\left(\frac{1}{e^{\beta(E_c(\textbf{k})-\mu)}+1}-\frac{1}{e^{\beta(E_v(\textbf{k})-\mu)}+1}\right)\\ \nonumber
		&=&\frac{\partial}{\partial_{k_x}}\left(\frac{1}{e^{\beta(E_c(\textbf{k})-\mu)}+1}-\frac{1}{e^{\beta(E_v(\textbf{k})-\mu)}+1}\right)\\ \nonumber
		&+&\frac{\partial}{\partial_{k_y}}\left(\frac{1}{e^{\beta(E_c(\textbf{k})-\mu)}+1}-\frac{1}{e^{\beta(E_v(\textbf{k})-\mu)}+1}\right)
	\end{eqnarray*}
	where, $ \beta=1/k_BT $, with $ k_B $ being the Boltzmann constant, $ T $ being the temperature and $ \mu $ being the chemical potential. Depending on the position of $ \mu $, we demonstrate the behavior of the quantities appearing in the above expression at a single $\textbf{k}=(k_x,k_y)$ point.
	\begin{itemize}
		\item If for a given $ \textbf{k} $, $ E_c(\textbf{k})>\mu $, \textit{i.e} $ E_c(\textbf{k})-\mu=x $
		\begin{eqnarray*}
			\lim_{\beta\rightarrow\infty}\frac{\partial}{\partial_{k_x}}\frac{1}{e^{\beta x}+1} &=& \lim_{\beta\rightarrow\infty}\frac{-\beta e^{\beta x}}{(e^{\beta x}+1)^2}\frac{\partial}{\partial_{k_x}}E_c(\textbf{k})\\
			&=&\lim_{\beta\rightarrow\infty}\frac{x}{2x^2e^{\beta x}}\\
			&=&0
		\end{eqnarray*}		
		Similarly, we can write
		\begin{equation}
			\frac{\partial}{\partial_{k_x}}\left(\frac{1}{e^{\beta(E_c(\textbf{k})-\mu)}+1}-\frac{1}{e^{\beta(E_v(\textbf{k})-\mu)}+1}\right) - 
		\end{equation}		
		$ 	\lim_{\beta\rightarrow\infty}\frac{\partial}{\partial_{k_x}}\frac{1}{e^{\beta(E_v(\textbf{k})-\mu)}+1} = 0 $ for $ E_v(\textbf{k})>\mu  $
		\item If for a given $ \textbf{k} $, $ E_c(\textbf{k})<\mu $ \textit{i.e} $ E_c(\textbf{k})-\mu=-x $
		\begin{eqnarray*}
			\lim_{\beta\rightarrow\infty}\frac{\partial}{\partial_{k_x}}\frac{1}{e^{-\beta x}+1} &=&  \lim_{\beta\rightarrow\infty}\frac{\beta e^{-\beta x}}{(e^{-\beta x}+1)^2}\frac{\partial}{\partial_{k_x}}E_c(\textbf{k})\\
			&=&0
		\end{eqnarray*}
		Similarly, we can show that $ 	\lim_{\beta\rightarrow\infty}\frac{\partial}{\partial_{k_x}}\frac{1}{e^{\beta(E_v(\textbf{k})-\mu)}+1} = 0 $ for $ E_v(\textbf{k})<\mu  $
		\item If for a given $ \textbf{k} $, $ E_c(\textbf{k})=\mu $ \textit{i.e} $ E_c(\textbf{k})-\mu=0 $
		\begin{eqnarray*}	
			\lim_{\beta\rightarrow\infty}\frac{\partial}{\partial_{k_x}}\frac{1}{e^{\beta(E_c(\textbf{k})-\mu)}+1} &=&\frac{\partial}{\partial_{k_x}}E_c(\textbf{k}) \lim_{\beta\rightarrow\infty}\frac{-\beta }{4}\\
			&=&\frac{\partial}{\partial_{k_x}}E_c(\textbf{k})\delta(E_c(\textbf{k})-\mu)\\
		\end{eqnarray*}
		Similarly, we have $\lim_{\beta\rightarrow\infty}\frac{\partial}{\partial_{k_x}}\frac{1}{e^{\beta(E_c(\textbf{k})-\mu)}+1} = \frac{\partial}{\partial_{k_x}}E_v(\textbf{k})\delta(E_v(\textbf{k})-\mu)$
	\end{itemize}
	Based on this we conclude that the term $ \bm{\nabla}_{\textbf{k}}(f^T_n(\textbf{k},\mu)-f^T_m(\textbf{k},\mu)) $ appearing in Eq.~\ref{eq:jfermi} has a contribution from only those $ \textbf{k} $ points at which either $ E_c(\textbf{k})=\mu  $ or $ E_v(\textbf{k})=\mu  $. Therefore the contribution appearing in the gradient term $ \bm{\nabla}_{\textbf{k}}\left(\frac{1}{e^{\beta(E_c(\textbf{k})-\mu)}+1}-\frac{1}{e^{\beta(E_v(\textbf{k})-\mu)}+1}\right) $ can be replaced by
	\begin{equation}
		\bm{\nabla}_{\textbf{k}} E_c(\textbf{k})\delta(E_c(\textbf{k})-\mu) - \bm{\nabla}_{\textbf{k}} E_v(\textbf{k})\delta(E_v(\textbf{k})-\mu)
	\end{equation}
	The variation of $ j^{\text{Fermi}} $ as function of $ \hbar\omega $ is shown in Fig.~\ref{fig:jfermi}. We see that as $ \omega\rightarrow 0 $, $ j^{\text{Fermi}}\rightarrow\infty. $ 
	
	\section{The effect of corrugation, doping-dependent Hartree interactions and ph-asymmetry on the one-photon absorption coefficient}
	\label{sec:appendix}
\begin{figure}
		\centering
		\includegraphics[scale=0.2]{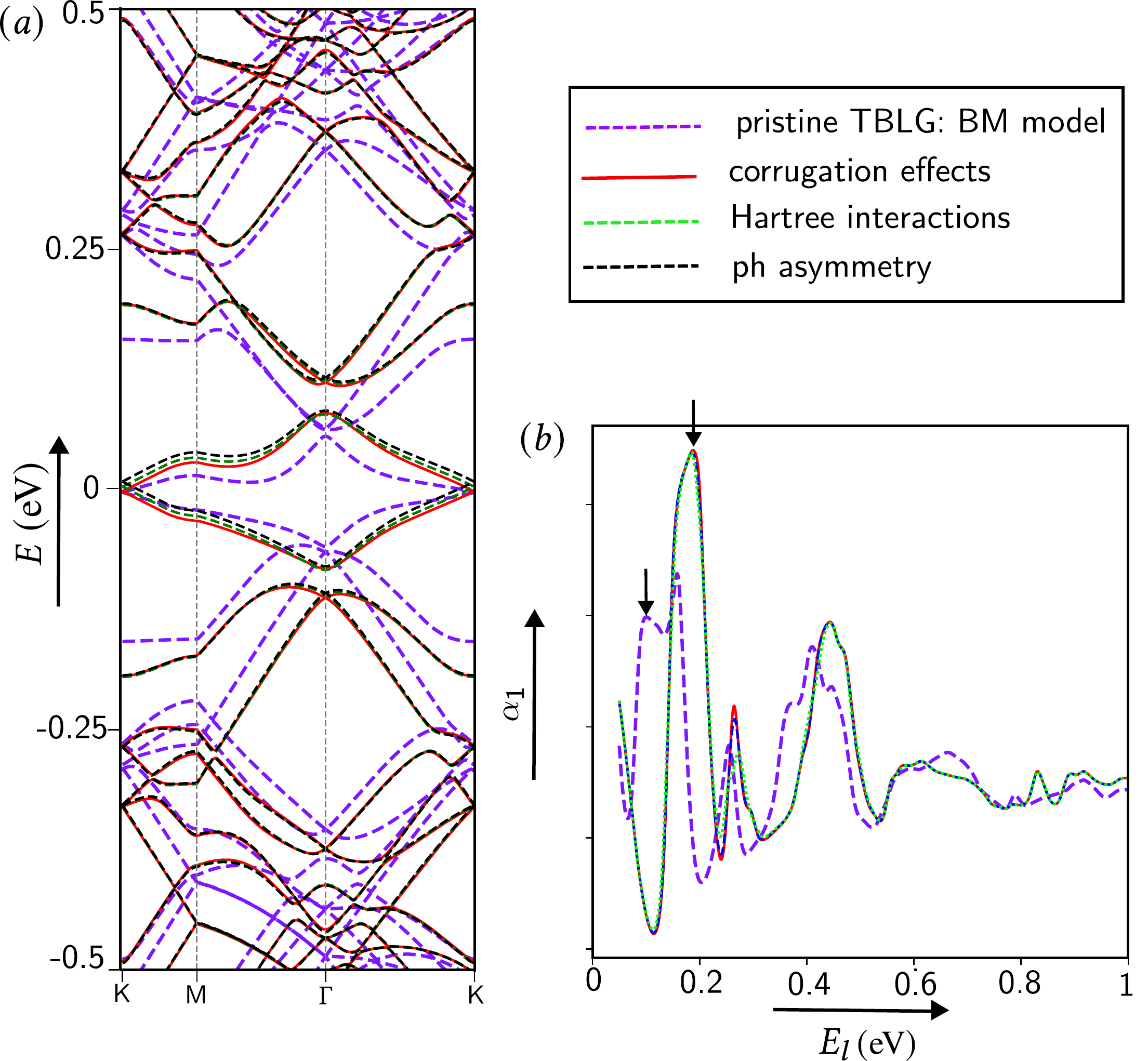}
		\caption{The (a) band structure and (b) behavior of the OPA coefficient $ \alpha_1 $ as a function of $ E_l $ for $ \theta = 1.54^{\circ} $. The legend mentions that the different colors indicate the step-wise modification added to the model.}
		\label{fig:app1}
	\end{figure}

	Bistritzer and MacDonald~\cite{Bistritzer2011} evaluated the band structure of pristine TBLG by setting $ w_0 = w_1 $. This model was later modified to include the effects of corrugation~\cite{Koshino2018}, doping-dependent Hartree interactions~\cite{Goodwin2020,Rademaker2019,Guinea2019,Guinea2022,Guinea2018} and particle-hole asymmetry~\cite{Rademaker2019,Vafek2023}. In FIG.~\ref{fig:app1}~(a), we show how the bands alter by including these modifications. The behavior of $ \alpha_1 $ for these four models is shown in FIG.~\ref{fig:app1}~(b). We infer that the resonant features of $ \alpha_1 $ (marked by the black arrows in FIG.~\ref{fig:app1}~(b)) are strongly altered by including the corrugation effects in pristine TBLG. As a result, the peak energy increases from $\sim 0.1~\si{\electronvolt}$ to $\sim 0.185~\si{\electronvolt}$. Any other modification in the system does not appreciably change the $ \alpha_1 $ behavior as a function of $ E_l $.

	\bibliography{ref.bib}
\end{document}